\documentclass[twocolumn]{aastex631}

\hypersetup{linkcolor=blue,colorlinks=true,citecolor=cyan,filecolor=black}

\received{06.06.2024}
\revised{16.08.2024}
\accepted{27.08.2024}

\submitjournal{ApJ}

\shorttitle{Improving Mid-Infrared Exo-Earth Retrievals with Physically Motivated Water Abundance Profiles and Cloud Models.}
\shortauthors{Konrad et al.}

\usepackage{graphicx}
\usepackage{placeins}
\usepackage{amsmath}
\usepackage{url}
\usepackage{hyperref} 
\usepackage{txfonts}
\usepackage{textgreek}
\usepackage{tabularx}
\usepackage{booktabs}
\usepackage{multirow}
\usepackage[version=4]{mhchem}
\usepackage{enumitem}
\usepackage{xcolor,colortbl}

\begin{document}

\title{Pursuing Truth: Improving Retrievals on Mid-Infrared Exo-Earth Spectra with Physically Motivated Water Abundance Profiles and Cloud Models.}

\correspondingauthor{Bj\"orn S. Konrad}
\email{konradb@student.ethz.ch}

\author[0000-0002-9912-8340]{Bj\"orn S. Konrad}
\affiliation{ETH Zurich, Institute for Particle Physics and Astrophysics, Wolfgang-Pauli-Strasse 27, CH-8093 Zurich, Switzerland}
\affiliation{National Center of Competence in Research PlanetS, Gesellschaftsstrasse 6, CH-3012 Bern, Switzerland}

\author[0000-0003-3829-7412]{Sascha P. Quanz}
\affiliation{ETH Zurich, Institute for Particle Physics and Astrophysics, Wolfgang-Pauli-Strasse 27, CH-8093 Zurich, Switzerland}
\affiliation{National Center of Competence in Research PlanetS, Gesellschaftsstrasse 6, CH-3012 Bern, Switzerland}
\affiliation{ETH Zurich, Department of Earth Sciences, Sonneggstrasse 5, 8092 Zurich, Switzerland}

\author[0000-0002-0006-1175]{Eleonora Alei}
\affiliation{NASA Postdoctoral Program Fellow, NASA Goddard Space Flight Center, Greenbelt, MD, USA}

\author[0000-0003-1127-8334]{Robin Wordsworth}
\affiliation{School of Engineering and Applied Sciences, Harvard University, Cambridge, MA 02138, USA}
\affiliation{Department of Earth and Planetary Sciences, Harvard University, Cambridge, MA 02138, USA}

\newcommand{\pt}[0]{\textit{P}$-$\textit{T}}

\newcommand{\R}[0]{\ensuremath{R}}
\newcommand{\Rv}[1]{\ensuremath{R = #1}}
\newcommand{\SN}[0]{\ensuremath{S/N}}
\newcommand{\SNv}[1]{\ensuremath{S/N = #1}}

\newcommand{\mic}[1]{\ensuremath{#1~\text{\textmu m}}}

\newcommand{\life}[0]{LIFE}
\newcommand{\lifesim}[0]{LIFE\textsc{sim}}

\newcommand{\Rpl}[0]{\ensuremath{R_{\text{pl}}}}
\newcommand{\Rearth}[0]{\ensuremath{R_\oplus}}   
\newcommand{\Mpl}[0]{\ensuremath{M_{\text{pl}}}}
\newcommand{\Teq}[0]{\ensuremath{T_\mathrm{eq}}}
\newcommand{\Teff}[0]{\ensuremath{T_\mathrm{eff}}}
\newcommand{\Ab}[0]{\ensuremath{A_\mathrm{B}}}
\newcommand{\Ps}[0]{\ensuremath{P_0}}
\newcommand{\Ts}[0]{\ensuremath{T_0}}
\newcommand{\Popaque}[0]{\ensuremath{P_\mathrm{Op.}}}
\newcommand{\Topaque}[0]{\ensuremath{T_\mathrm{Op.}}}
\newcommand{\Tct}[0]{\ensuremath{T_\mathrm{cloud-top}}}

\newcommand{\Pv}[0]{\ensuremath{p_\mathrm{v}}}
\newcommand{\Pvwater}[0]{\ensuremath{p_\mathrm{v,\,water}}}
\newcommand{\Pvice}[0]{\ensuremath{p_\mathrm{v,\,ice}}}
\newcommand{\PPwater}[0]{\ensuremath{p_\mathrm{\ce{H2O}}}}
\newcommand{\Pcloud}[0]{\ensuremath{p_\mathrm{C}}}
\newcommand{\VMRwater}[0]{\ensuremath{\nu_\mathrm{\ce{H2O}}}}
\newcommand{\VMRwaterDry}[0]{\ensuremath{\nu_\mathrm{\ce{H2O},\,dry}}}
\newcommand{\drying}[0]{\ensuremath{\delta_\mathrm{H_2O}}}
\newcommand{\fcloud}[0]{\ensuremath{f_\mathrm{C}}}

\newcommand{\modelparam}[0]{\ensuremath{\boldsymbol{\checkmark}}}
\newcommand{\notparam}[0]{\ensuremath{\boldsymbol{\times}}}

\newcommand{\cwcf}[0]{\ensuremath{\mathcal{M}_\mathrm{C,CF}}}
\newcommand{\vwcf}[0]{\ensuremath{\mathcal{M}_\mathrm{V,CF}}}
\newcommand{\vwcl}[0]{\ensuremath{\mathcal{M}_\mathrm{V,CL}}}

\newcommand{\lgrt}[1]{\ensuremath{\log_{10}(#1)}}
\newcommand{\lgrtdaj}[1]{\ensuremath{\log_{10}\left(#1\right)}}
\newcommand{\lgrtshort}[1]{\ensuremath{L\left(#1\right)}}

\definecolor{tab:cwcf}{RGB}{213,  94, 0}
\definecolor{tab:vwcf}{RGB}{0, 158, 115}
\definecolor{tab:vwcl}{RGB}{0, 114, 178}
\begin{abstract}
    Atmospheric retrievals are widely used to constrain exoplanet properties from observed spectra. We investigate how the common nonphysical retrieval assumptions of vertically constant molecule abundances and cloud-free atmospheres affect our characterization of an exo-Earth {(an Earth-twin orbiting a Sun-like star)}. Specifically, we use a state-of-the-art retrieval framework to explore how assumptions for the \ce{H2O} profile and clouds affect retrievals. In a first step, we validate different retrieval models on a low-noise simulated 1D mid-infrared (MIR) spectrum of Earth. Thereafter, we study how these assumptions affect the characterization of Earth with the Large Interferometer For Exoplanets (\life{}). We run retrievals on \life{} mock observations based on real disk-integrated MIR Earth spectra. The performance of different retrieval models is benchmarked against ground truths derived from remote sensing data. We show that assumptions for the \ce{H2O} abundance and clouds directly affect our characterization. Overall, retrievals that use physically motivated models for the \ce{H2O} profile and clouds perform better on the empirical Earth data. For observations of Earth with \life{}, they yield accurate estimates for the radius, pressure-temperature structure, and the abundances of \ce{CO2}, \ce{H2O}, and \ce{O3}. Further, at \Rv{100}, a reliable and bias-free detection of the biosignature \ce{CH4} becomes feasible. We conclude that the community must use a diverse range of models for temperate exoplanet atmospheres to build an understanding of how different retrieval assumptions can affect the interpretation of exoplanet spectra. This will enable the characterization of distant habitable worlds and the search for life with future space-based instruments.
\end{abstract}
\section{Introduction}\label{sec:introduction}

    Both the Astrobiology Strategy \citep{Astrobiology_Strategy_2017} and the Astro 2020 Decadal Survey in the United States \citep{Decadal_Study_2021} identify the atmospheric characterization of terrestrial exoplanets as a key endeavor for exoplanet science. The measurement and subsequent analysis of an exoplanet's spectrum allows us to constrain key atmospheric properties such as the pressure-temperature (\pt{}) structure and the molecular composition. Such constraints shed light on the planet's habitability and could enable the detection of biological activity.
    
    Temperate exoplanets with Earth-like radii and masses that reside within their host star's habitable zone \citep[HZ;][]{HZ_Kasting93,HZ_kopparapu13} are of particular interest. Findings from transit surveys such as the Kepler mission \citep{Borucki2010} or the Transiting Exoplanet Survey Satellite \citep[TESS;][]{TESS_2015} and current long-term radial velocity (RV) surveys predict that these planets are a common occurrence \citep[e.g.,][]{Petigura2013, F&M2014, DressingCharbonneau2015, Bryson2021}. Several rocky HZ exoplanets have been found within 20~pc of the sun \citep[see, e.g.,][for a catalogue]{Hill_2023} via both the transit \citep[e.g.,][]{Bthompson2015, Gillon2017, Vanderspek2019} and the RV \citep[e.g.,][]{Anglada2016, Ribas2016, Zechmeister2019} methods. Currently, transit observations with the James Webb Space Telescope (JWST) are investigating whether rocky exoplanets orbiting M dwarf stars can retain significant atmospheres \citep[e.g.,][]{Koll2019, Greene2023, Zieba2023, Lustig2023, Ih2023, Lincowski2023, Madhusudhan2023, Lim2023}. Yet, the JWST will likely not provide a detailed atmospheric characterization of these objects \citep[e.g.,][]{morley2017, Krissansen-Totton2018_Biosigs_JWST}. In the near future, ground-based extremely large telescopes (ELTs) will directly measure the reflected stellar light and thermal emission of the closest HZ exoplanets \citep[e.g.,][]{quanz2015, Bowens2021, Kasper2021}. However, a detailed atmospheric characterization for a statistically significant number (dozens) of rocky HZ exoplanets is not achievable by any current or approved future instrument.
    
    Thus, the exoplanet community is pushing for a next generation of space-based observatories. Motivated by the LUVOIR \citep{LUVOIR_2019} and HabEx \citep{Gaudi2020} mission concepts, The Astro 2020 Decadal Survey in the United States \citep{Decadal_Study_2021} recommended the space-based Habitable Worlds Observatory (HWO). HWO aims to directly detect host-star light scattered by rocky HZ exoplanets in the ultraviolet, optical, and near-infrared (UV/O/NIR) wavelength range. Yet, an exoplanet's mid-infrared (MIR) thermal emission spectrum also encodes unique information about the atmospheric state and the surface conditions \citep[e.g.,][]{DesMarais2002, Hearty2009, Catling2018, Schwieterman2018, mettler_2020, mettler_2023}. 
    {Thus, the direct detection of the thermal emission of rocky, temperate exoplanets in the MIR has been identified as one of the most important science topics to be considered for the future science program of the European Space Agency (ESA) \citep{ESAV2050}}. This was triggered by a White Paper submitted by members of the Large Interferometer For Exoplanets (\life{}) initiative \citep[] {Quanz:exoplanets_and_atmospheric_characterization}, which is working on developing a large space-based MIR nulling interferometer \citep[][]{K&QLIFE, LIFE_I}.
    

    Several studies have investigated the \life{} instrument design \citep{LIFE_II, LIFE_IV, LIFE_VII, LIFE_XI} and its potential to detect rocky HZ exoplanets \citep{LIFE_I, LIFE_VI, LIFE_X}. Further studies focused on \life{}'s potential to characterize terrestrial exoplanets by studying simulated 1D spectra. \citet{LIFE_III} derived first constraints for the wavelength coverage, {spectral resolving power (\R{})}, and signal-to-noise (\SN{}) levels by studying a cloud-free modern Earth. Subsequent studies reevaluated these requirements by analyzing Earth at different epochs of its evolution \citep{LIFE_V}, a cloudy Venus \citep{LIFE_IX}, and the detectability of biosignature gases \citep{LIFE_VIII, LIFE_XII}. {Last, \citet{Alei2024arX} investigated how combined observations of an Earth-twin with HWO and LIFE could improve its characterization. Thereby, the uniqueness of the information attainable from a planet's MIR emission is highlighted.}

    These \life{} studies all analyze spectra that were generated using a 1D atmosphere model. However, \citet{mettler_2020} find significant discrepancies between thermal emission spectra from different locations on Earth. In \citet{mettler_2023}, real disk-integrated rather than local Earth MIR emission spectra are analyzed. The authors find that Earth’s MIR emission depends significantly on the observed viewing geometry and season. Thus, a representative, disk-integrated MIR spectrum does not exist for Earth. In a third study, \citet{Mettler2024} treat Earth as an exoplanet observed with \life{}. By running atmospheric retrievals on empirical MIR disk-integrated spectra from Earth observing satellites, the authors study how their characterization of Earth depends on the viewing angle and season. They conclude that viewing geometry and season only minorly impact their results and successfully characterize Earth as a temperate habitable planet, featuring biosignature gases.

    However, when comparing the retrieved parameter estimates with remote sensing ground truth data, \citet{Mettler2024} observe significant offsets between the two. The authors predominantly attribute these biases to two simplifying retrieval assumptions. First, their retrieval model assumes all abundance profiles to be constant throughout the atmosphere. Especially for the strong MIR absorber \ce{H2O}, whose abundance varies by several orders of magnitude throughout Earth's atmosphere, this is a strong simplification. Second, they neglect Earth's patchy cloud coverage by assuming a cloud-free atmosphere in all retrieval models.
    
    The main aim of this study is to demonstrate that simple physical models for the \ce{H2O} structure and the clouds in Earth's atmosphere can significantly reduce such biases. First, we validate our models with retrievals on simulated 1D Earth spectra. Later, we run retrievals on the same empirical disk-integrated Earth spectra considered in \citet{Mettler2024} and compare the results with ground truth remote sensing data. Such studies on real observations of terrestrial solar system planets are uncommon \citep[e.g.,][]{Tinetti_2006_1, Robinson_2023, Lustig-Yaeger_2023}. Yet, they provide a powerful approach to help us understand how common retrieval assumptions can bias our results. Building this understanding is indispensable to ensure the correct characterization of terrestrial HZ exoplanets in the future.
\section{Datasets and Methods}\label{sec:methods}

    Here, we introduce our retrieval routine and the datasets studied. In Section~\ref{subsec:retrieval_routine}, we provide information on the retrieval routine and our models for the \ce{H2O} abundance profile and clouds. Next, in Section~\ref{subsec:atmospheric_models}, we introduce the three different retrieval forward models considered. Last, we introduce the different input spectra and noise models, and justify the assumed priors in Sections~\ref{subsec:retrieval_inputs} and \ref{subsec:priors}.

\subsection{Atmospheric retrieval routine}\label{subsec:retrieval_routine}

    We run retrievals using a modified version of the retrieval routine initially introduced in \citet{LIFE_III} and later improved and modified in \citet{LIFE_V} and \citet{LIFE_IX}. Here, we provide a brief summary of the original routine and focus on the additions to the routine made for this publication. For an in-depth description of the retrieval framework, we refer to the original publications.

    The retrieval framework is based on the radiative transfer code \texttt{petitRADTRANS} \citep{Molliere:petitRADTRANS, Molliere:petitRADTRANS2, LIFE_V}, which is used to calculate the theoretical MIR thermal emission spectrum of a 1D plane-parallel parametric atmosphere model. The model atmosphere is defined via a set of forward model parameters (see Section~\ref{subsec:atmospheric_models} for a description of the forward models we used). To calculate the MIR emission spectrum corresponding to a set of model parameters, \texttt{petitRADTRANS} assumes the planetary surface to emit black-body radiation. It then models the interaction of each discrete atmospheric layer with the radiation by accounting for absorption, emission, and scattering. This yields the MIR flux at the top of the atmosphere.
    
    The goal of a retrieval is to search the space spanned by the forward model parameters for combinations of parameter values that best reproduce an observed planet spectrum. The prior probability distributions (or ``priors") of the model parameters specify the parameter space to be probed by the retrieval. To efficiently explore this prior space, we use \texttt{py}-\texttt{MultiNest} \citep{Buchner:PyMultinest}, a python package based on the \texttt{MultiNest} \citep{Feroz:Multinest} implementation of Nested Sampling \citep{Skilling:Nested_Sampling}. For this study, we ran all retrievals using 700 live points and a sampling efficiency of 0.3\footnote{As suggested for evidence evaluation by the \texttt{MultiNest} documentation: \url{https://github.com/farhanferoz/MultiNest}.}.

    The output of a retrieval are posterior probability distributions (or ``posteriors") for the model parameters. The posteriors summarize how likely different combinations of parameter values are. In addition, our retrieval routine yields estimates of the Bayesian evidence $\mathcal{Z}$, which quantifies how well the forward model fits the input spectrum. Thus, $\mathcal{Z}$ provides a means of comparing the performance of different forward models. If we run two retrievals on the same spectrum using different forward models $\mathcal{A}$ and $\mathcal{B}$, the retrievals will yield different log-evidences $\ln\left(\mathcal{Z_A}\right)$ and $\ln\left(\mathcal{Z_B}\right)$. From the log-evidences, we calculate the Bayes' factor $K$:
    \begin{equation}\label{eq:BayesFactor}
        \lgrt{K}=\frac{\ln\left(\mathcal{Z}_{\mathcal{A}}\right)-\ln\left(\mathcal{Z}_{\mathcal{B}}\right)}{\ln\left(10\right)}.
    \end{equation}
    The value of $K$ specifies which model is preferred. A positive $\lgrt{K}$ marks a preference for model $\mathcal{A}$, whereas negative values indicate preference for $\mathcal{B}$. The Jeffreys scale \citep[][Table~\ref{Table:Jeffrey}]{Jeffreys:Theory_of_prob} provides a quantification for the strength of the model preference.

    \begin{deluxetable}{lcc}
\tablecaption{Jeffreys scale \citep{Jeffreys:Theory_of_prob}.}
\label{Table:Jeffrey}      
\tablehead{
\colhead{$\lgrt{K}$} &\colhead{Probability} &\colhead{Strength of Evidence}
}
\startdata 
   $<0$     &$<0.5$         &Support for $\mathcal{B}$\\
   $0-0.5$  &$0.5-0.75$     &Very weak support for $\mathcal{A}$\\
   $0.5-1$  &$0.75-0.91$    &Substantial support for $\mathcal{A}$\\
   $1-2$    &$0.91-0.99$    &Strong support for $\mathcal{A}$\\
   $>2$     &$>0.99$        &Decisive support for $\mathcal{A}$\\ 
\enddata
\tablecomments{Scale to interpret the Bayes' factor $K$ for two models $\mathcal{A}$ and $\mathcal{B}$. The scale is symmetrical, i.e., negative values of $\lgrt{K}$ correspond to very weak, substantial, strong, or decisive support for model $\mathcal{B}$.}
\end{deluxetable}

\subsubsection{Model for water condensation}\label{subsec:Water_Cond}

\begin{figure}
  \centering
    \includegraphics[width=0.47\textwidth]{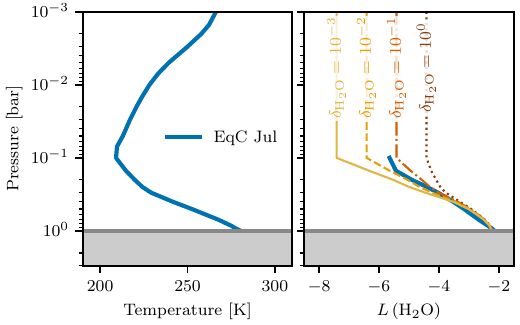}
  \caption{\ce{H2O} profiles calculated using the condensation model (Section~\ref{subsec:Water_Cond}) for the EqC~Jul Earth view (Section~\ref{subsubsec:input_real_earth}). \textit{Left panel:} The ground truth \pt{} profile for the EqC~Jul view of Earth. \textit{Right panel:} The ground truth \ce{H2O} profile as a thick blue line together with the \ce{H2O} profiles calculated for different \drying{} values.}
  \label{fig:cond_model}
\end{figure}

    Retrievals commonly assume that the atmospheric trace-gas abundances are vertically constant. While this assumption helps reduce the computational complexity, recent studies have shown that it can bias the retrieved posteriors \citep[e.g.,][]{Rowland2023}. For Earth, \ce{H2O} has strong spectral features throughout the MIR \citep[e.g., Figure~1 in][]{LIFE_III}. The bulk of Earth's MIR emission originates from the lowermost atmospheric layers (between 1~bar and 0.1~bar), {where the atmospheric \ce{H2O} content decreases strongly with altitude by more than 3~orders of magnitude \citep[see, e.g., Figure~\ref{fig:cond_model};][]{Mettler2024}}. The main process responsible for this decrease in \ce{H2O} in Earth's lower atmosphere is the convection of moist air from the surface to higher atmospheric layers. The rising moist air cools causing \ce{H2O} to condense, form clouds, and rain out. Consequentially, the \ce{H2O} abundance drops significantly with altitude in Earth's atmosphere, and we expect signatures thereof to be imprinted in the MIR emission.

    To model the condensation induced vertical variations in the \ce{H2O} profile in our retrieval forward model, we start by calculating the saturation vapor pressure of \ce{H2O} for each atmospheric layer. For this, we use the temperature $T$ of the layer as well as one of the two experimentally determined Goff-Gratch equations \citep[Eqs.~\ref{eq:P_sat_water} and \ref{eq:P_sat_ice};][]{Goff1946,Goff1957}. If $T$ is greater than the triple point temperature of \ce{H2O} ($T_\mathrm{tp}=273.16$~K), we use the equation for the saturation vapor pressure over water (\Pvwater{} in hPa):
    \begin{equation}\label{eq:P_sat_water}
        \begin{split}
            \Pvwater\left(T\right)&=l_1\cdot(1-T_\mathrm{st}/T)+ l_2\cdot\lgrt{T_\mathrm{st}/T}\\
            &+l_3\cdot10^{-7}\cdot\left(1-10^{l_4\cdot\left(1-T/T_\mathrm{st}\right)}\right)\\
            &+l_5\cdot10^{-3}\cdot\left(10^{l_6\cdot\left(1-T_\mathrm{st}/T\right)}-1\right)+\lgrt{p_\mathrm{st}}.
        \end{split}
    \end{equation}
    Here, $T_\mathrm{st}=373.15$~K and $p_\mathrm{st}=1013.25$~hPa are the steam-point temperature and pressure of \ce{H2O}, and the factors $l_i$ are constants: $l_1=7.90298$, $l_2=5.02808$, $l_3=1.3816$, $l_4=11.344$, $l_5=8.1328$, and $l_6=3.49149$. Conversely, if $T$ lies below $T_\mathrm{tp}$, we use the Goff-Gratch equation for the saturation vapor pressure over ice (\Pvice{} in hPa):
    \begin{equation}\label{eq:P_sat_ice}
        \begin{split}
            \Pvice\left(T\right)&=s_1\cdot(1-T_\mathrm{tp}/T)-s_2\cdot\lgrt{T_\mathrm{tp}/T}\\
            &+s_3\cdot(1-T/T_\mathrm{tp})+\lgrt{p_\mathrm{tp}}.
        \end{split}
    \end{equation}
    Here, $p_\mathrm{tp}=6.1173$~hPa is the triple point pressure of \ce{H2O}, and the factors $s_i$ are constants: $s_1=9.09718$, $s_2=3.56654$, and $s_3=0.876793$. The vapor pressure of \ce{H2O} (\Pv{}) as a function of $T$ can be summarized as:
    \begin{equation}
        \Pv\left(T\right)= 
        \begin{cases}
            \Pvwater\left(T\right),& \text{if } T\geq T_\mathrm{tp},\\
            \Pvice\left(T\right),  & \text{otherwise}.
        \end{cases}
    \end{equation}
    Next, we assume a vertically constant volume mixing ratio of \ce{H2O} (\VMRwater{}). Starting from the surface layer, we follow the atmospheric \pt{} structure to the higher atmospheric layers with pressures $p$ and temperatures $T$. For each layer, we compare the partial pressure of \ce{H2O} ($\PPwater{}=\VMRwater\cdot p$) to $\Pv(T)$. If \PPwater{} lies below $\Pv(T)$, we proceed to the next layer. In contrast, if \PPwater{} in a layer exceeds $\Pv(T)$, we reduce \VMRwater{} in that layer and all lower-pressure layers such that \PPwater{} in the current layer equals $\Pv(T)$. This yields a $\VMRwater{}(p)$ profile that is either constant or decreasing with altitude (depending on the \pt{} structure and \VMRwater{} in the surface layer).

    This condensation model accurately estimates $\VMRwater{}(p)$ in Earth's lower troposphere ($p>0.3$~bar), where \PPwater{} is comparable to $\Pv{}(T)$. However, it significantly overestimates the $\VMRwater{}(p)$ in Earth's dry upper troposphere and stratosphere ($p<0.3$~bar), where \PPwater{} is significantly smaller than $\Pv(T)$. In our retrievals, we allow for a dry upper atmosphere by adding a drying parameter, $\drying{}\in\left[0,1\right]$, to the forward model. Using the parameter \drying{}, we calculate the dried \ce{H2O} profile, $\VMRwaterDry(p)$, as follows:
    \begin{equation}\label{eq:H2O_Drying}
        \VMRwaterDry(p)=\VMRwater(p)\cdot\drying^{\VMRwater{}_\mathrm{,\,TOA}/\VMRwater(p)}.
    \end{equation}
    Here, $\VMRwater{}_\mathrm{,\,TOA}$ is $\VMRwater(p)$ in the uppermost atmospheric layer. Equation~\ref{eq:H2O_Drying} is designed such that the drying is strongest for small values of \drying{} and for atmospheric layers where $\VMRwater(p)$ is comparable to $\VMRwater{}_\mathrm{,\,TOA}$. A value of $\drying{}=1$ results in $\VMRwaterDry(p)=\VMRwater(p)$, which means that there is no drying in the upper atmosphere.
    
    In Figure~\ref{fig:cond_model}, we demonstrate that our condensation model yields realistic estimates for Earth's \ce{H2O} profile. We compare the ground truth \ce{H2O} profile for the disk-integrated average equatorial view of Earth in July (EqC~Jul) with \ce{H2O} profiles that we calculated from the EqC~Jul \pt{} structure using our condensation model and different \drying{} values (see Section~\ref{subsubsec:input_real_earth} for information on the EqC~Jul view). {We observe that our condensation model successfully approximates Earth's true \ce{H2O} profile for $\drying{}=10^{-1}-10^{-2}$. The value of \drying{} mainly affects the \ce{H2O} profile for $p<0.3$~bar and is required to accurately model Earth's \ce{H2O} structure. If drying is neglected (brown-dotted line; $\drying{}=10^0$), we fail to accurately reproduce Earth's \ce{H2O} profile.}

\subsubsection{Model for partial cloud coverage}\label{subsec:Cloud_Model}

    The condensation of \ce{H2O} in Earth's atmosphere is linked to the formation of \ce{H2O} clouds. Thus, if \ce{H2O} condenses in our model atmosphere, we expect clouds to form. To account for this effect, we add the option to include a simple model for partial cloud coverage in the retrieval forward model.

    {To account for clouds in our retrievals, we position gray clouds }
    {at an atmospheric pressure of \Pcloud{}. We calculate \Pcloud{} by taking the median pressure of all atmosphere layers where \ce{H2O} condensation occurs that exhibit a relative humidity greater than 10\% (i.e., $\PPwater(p)/\Pv(T)>0.1$). This assumption for \Pcloud{} positions the clouds within the atmospheric layers where \ce{H2O} condenses. If no \ce{H2O} condensation occurs, we assume the atmosphere to be cloud-free.}
    
    To model an Earth-like partial cloud cover, we use our retrieval forward model to calculate the spectra for a cloud-free ($F_\mathrm{CF}$) and cloudy ($F_\mathrm{C}$) model atmosphere. We then mix the two spectra as follows to obtain the total flux $F_\mathrm{tot}$ as a function of the wavelength $\lambda$:
    \begin{equation}
        F_\mathrm{tot}(\lambda) = \fcloud\cdot F_\mathrm{C}(\lambda) + \left(1-\fcloud\right)\cdot F_\mathrm{CF}(\lambda).
    \end{equation}
    Here, the forward model parameter $\fcloud{}\in\left[0,1\right]$ describes the cloud-fraction. A value of $\fcloud{}=0$ corresponds to the cloud-free case, whereas $\fcloud{}=1$ yields a fully cloudy atmosphere.

\subsection{Retrieval forward models}\label{subsec:atmospheric_models}
\begin{table*}
\centering
\caption{Summary of the three different retrieval forward models.}
\label{tab:model_summary} 
\begin{tabular}{lcl}
\hline\hline
Model &Color-coding &Model description\\\hline
\cwcf{} &\cellcolor{tab:cwcf}\qquad &Vertically constant \ce{H2O} abundance, cloud-free\\
\vwcf{} &\cellcolor{tab:vwcf}\qquad &\ce{H2O} abundance from condensation model, cloud-free\\
\vwcl{} &\cellcolor{tab:vwcl}\qquad &\ce{H2O} abundance from condensation model, partial cloud coverage\\
\hline
\end{tabular}
\tablecomments{The second column shows the color-coding used for the models throughout this publication.}
\end{table*}
\begin{deluxetable*}{lcccccc}
\tablecaption{Model parameters, values for the validation retrieval, priors, and model configurations.}
\label{tab:model_parameters} 
\tablehead{
\colhead{\multirow{2}{*}{Parameter}} &\colhead{\multirow{2}{*}{Description}} &\colhead{\multirow{2}{*}{\begin{tabular}{c}Validation\\True Values\end{tabular}}} &\colhead{\multirow{2}{*}{Prior}} &\multicolumn{3}{c}{Model Configuration}
\\
\cline{5-7}
&&&&\colhead{\cwcf{}} &\colhead{\vwcf{}} &\colhead{\vwcl{}}}
\startdata 
 $a_4$               &\pt{} parameter (degree 4)                         & 1.69  &$\mathcal{U}(0,10)$  &\modelparam{} &\modelparam{} &\modelparam{}\\
 $a_3$               &\pt{} parameter (degree 3)                         & 23.12  &$\mathcal{U}(0,100)$    
&\modelparam{} &\modelparam{} &\modelparam{}\\
 $a_2$               &\pt{} parameter (degree 2)                         & 99.70  &$\mathcal{U}(0,500)$ 
        &\modelparam{} &\modelparam{} &\modelparam{}\\
 $a_1$               &\pt{} parameter (degree 1)                         & 146.63  &$\mathcal{U}(0,500)$    
        &\modelparam{} &\modelparam{} &\modelparam{}\\
 $a_0$               &\pt{} parameter (degree 0)                         & 285.22   &$\mathcal{U}(0,1000)$  
        &\modelparam{} &\modelparam{} &\modelparam{}\\
 $\lgrtshort{\Ps{}}$      &$\lgrtshort{\textrm{Surface pressure }\left[\mathrm{bar}\right]}$ & 0.01  &$\mathcal{U}(-4,2)$   
        &\modelparam{} &\modelparam{} &\modelparam{}\\
 $\Rpl{}$            &Planet radius $\left[R_\oplus\right]$              & 1.00  &$\mathcal{G}(1.0,0.2)$ 
        &\modelparam{} &\modelparam{} &\modelparam{}\\
 $\lgrtshort{\Mpl{}}$     &$\lgrtshort{\textrm{Planet mass } \left[M_\oplus\right]}$ & 0.00  &$\mathcal{G}(0.0,0.4)$ 
        &\modelparam{} &\modelparam{} &\modelparam{}\\
 $\lgrtshort{\ce{N2}}$    &$\lgrtshort{\textrm{\ce{N2} mass fraction}}$            & -0.10 &$\mathcal{U}(-10,0)$ 
        &\modelparam{} &\modelparam{} &\modelparam{}\\
 $\lgrtshort{\ce{O2}}$    &$\lgrtshort{\textrm{\ce{O2} mass fraction}}$            & -0.68 &$\mathcal{U}(-10,0)$ 
        &\modelparam{} &\modelparam{} &\modelparam{}\\
 $\lgrtshort{\ce{CO2}}$   &$\lgrtshort{\textrm{\ce{CO2} mass fraction}}$           & -3.39 &$\mathcal{U}(-10,0)$ 
        &\modelparam{} &\modelparam{} &\modelparam{}\\
 $\lgrtshort{\ce{O3}}$    &$\lgrtshort{\textrm{\ce{O3} mass fraction}}$           & -6.52 &$\mathcal{U}(-10,0)$ 
        &\modelparam{} &\modelparam &\modelparam{}\\
 $\lgrtshort{\ce{CH4}}$   &$\lgrtshort{\textrm{\ce{CH4} mass fraction}}$           & -5.77  &$\mathcal{U}(-10,0)$ 
        &\modelparam{} &\modelparam{} &\modelparam{}\\
 $\lgrtshort{\ce{H2O}}$   &$\lgrtshort{\textrm{\ce{H2O} mass fraction}}$            & -2.10 &$\mathcal{U}(-10,-1)$ 
        &\modelparam{}   &\modelparam{} &\modelparam{}\\
 $\lgrtshort{\drying}$   &$\lgrtshort{\textrm{\ce{H2O} drying parameter}}$          & -2.00  &$\mathcal{U}(-5,0)$ 
        &\notparam{}   &\modelparam{} &\modelparam{}\\
 \fcloud{}   &Cloud fraction        &  0.67   &$\mathcal{U}(0,1)$
        &\notparam{}   &\notparam{} &\modelparam{}\\
\enddata
\tablecomments{Here, $L(\cdot)$ stands for $\lgrt{\cdot}$. In the third column we specify the values assumed to generate the mock-Earth spectrum for the validation retrievals. The listed \ce{H2O} value is the abundance at the surface obtained from the condensation model. The \fcloud{} value is motivated by the average cloud fractions presented in \citet{Mettler2024} and \drying{} is chosen such that the calculated \ce{H2O} profile fits Earth's profile. All other parameter values are taken from \citet{LIFE_III}. In the fourth column, the assumed priors are listed. We denote a boxcar prior with lower threshold $x$ and upper threshold $y$ as $\mathcal{U}(x,y)$; For a Gaussian prior with mean $\mu$ and standard deviation $\sigma$, we write $\mathcal{G}(\mu,\sigma)$. The last three columns summarize the model parameters used by the different retrieval forward models (\modelparam{}~$=$~used, \notparam{}~$=$~unused).}
\end{deluxetable*}

    We run retrievals using the three different forward models listed in Table~\ref{tab:model_summary}. A complete list of the parameters used by each model is provided in Table~\ref{tab:model_parameters}. All models assume a 1D atmosphere consisting of 100 equally thick layers (in log-space) between $10^{-4}$~bar and the surface pressure \Ps{}. Each individual layer is characterized by its pressure $P$, the corresponding temperature $T$, and the opacity sources present. As in \citet{LIFE_III,LIFE_IX}, \citet{LIFE_V,Alei2024arX}, and \citet{Mettler2024}, we parameterize the atmospheric \pt{} structure by means of a fourth order polynomial:
    \begin{equation}\label{equ:4poly}
        T(P)=\sum_{i=0}^4a_iP^i.
    \end{equation}
    The factors $a_i$ are the five forward model parameters describing the atmospheric \pt{} structure. This choice is motivated by \citet{LIFE_III}, who demonstrate that a polynomial \pt{} model helps reduce a retrieval's computational complexity by minimizing the number of \pt{} parameters\footnote{In a recent publication, \citet{TimmyPT2023} demonstrated that learning-based \pt{} models can reduce the number of \pt{} parameters further. However, the accuracy of such models for terrestrial \pt{} structures is currently limited by the availability of sufficient training data.}.

    Further, all models assume the same spectroscopically active molecules to be present. We model MIR molecular absorption and emission features of \ce{CO2}, \ce{H2O}, \ce{O3}, and \ce{CH4}\footnote{This choice of molecules is motivated by a detailed Bayesian model comparison study performed by \citet{Mettler2024} for the same disk-integrated Earth spectra.}. The used line lists, broadening coefficients, and cutoffs are summarized in Table~\ref{tab:opacities}. We also consider collision-induced absorption (CIA) and Rayleigh scattering features of atmospheric molecules (all considered CIA-pairs and Rayleigh-species are summarized in Table~\ref{tab:opacities}). Further, we use a spectroscopically inactive filling gas with the mean molecular weight of \ce{N2} to ensure that: $\sum(\mathrm{gas\,abundances})=1$. Last, to determine the scale height of the model atmosphere, we require the planetary surface gravity, which we calculate from the radius and mass parameters (\Rpl{}, \Mpl{}).


    The three forward models only differ in the assumptions made for the \ce{H2O} abundance profile and the clouds. The simplest model (\cwcf{}) assumes constant abundance profiles for all molecules and a cloud-free atmosphere. This model is equivalent to the forward model used in \citet{Mettler2024}. The intermediate complexity model (\vwcf{}) again assumes the atmosphere to be cloud-free. However, it allows for a non-constant \ce{H2O} abundance profile by modeling \ce{H2O} condensation as described in Section~\ref{subsec:Water_Cond}. The most complex of the three models (\vwcl{}) also accounts for \ce{H2O} condensation. In addition, it relaxes the assumption of a cloud-free atmosphere by modeling partial cloud coverage following the method outlined in Section~\ref{subsec:Cloud_Model}.

\begin{deluxetable*}{cccccccccc}
\tablecaption{Line and continuum opacities used for this study.}
\label{tab:opacities}
\tablehead{\multicolumn{4}{c}{Molecular Line Opacities}&&\multicolumn{2}{c}{CIA}&&\multicolumn{2}{c}{Rayleigh Scattering}\\\cline{1-4}\cline{6-7}\cline{9-10}
\colhead{Molecule} & \colhead{Line List}& \colhead{Pressure-broadening}& \colhead{Wing cutoff}&&\colhead{Pair}&\colhead{Reference}&&\colhead{Molecule}&\colhead{Reference}
}
\startdata
\ce{CO2}    &HN20     &$\gamma_{\mathrm{air}}$    &25 cm$^{-1}$ &&\ce{N2}$-$\ce{N2}     &KA19           &&\ce{N2}    &TH14, TH17\\
\ce{H2O}    &HN20     &$\gamma_{\mathrm{air}}$    &25 cm$^{-1}$ &&\ce{N2}$-$\ce{O2}     &KA19           &&\ce{O2}    &TH14, TH17\\
\ce{O3}     &HN20     &$\gamma_{\mathrm{air}}$    &25 cm$^{-1}$ &&\ce{O2}$-$\ce{O2}     &KA19           &&\ce{CO2}   &SU05\\
\ce{CH4}    &HN20     &$\gamma_{\mathrm{air}}$    &25 cm$^{-1}$ &&\ce{CO2}$-$\ce{CO2}   &KA19           &&\ce{CH4}   &SU05\\
... & ... & ... & ... &&\ce{CH4}$-$\ce{CH4}   &KA19   && \ce{H2O}   &HA98\\
... & ... & ... & ... &&\ce{H2O}$-$\ce{H2O}   &KO21   && ...        &... \\
... & ... & ... & ... &&\ce{H2O}$-$\ce{N2}    &KO21   && ...        &... \\
\enddata
\tablerefs{(HA98) \citet{Harvey1998}; (HN20) \citet{HN20}; (KA19) \citet{KARMAN2019160}; (KO21) \citet{Kofman2021}; (SU05) \citet{2005JQSRT..92..293S}; (TH14) \citet{TH14}; (TH17) \citet{TH17}.}
\end{deluxetable*}

\subsection{Input spectra for the retrievals}\label{subsec:retrieval_inputs}

    {In the context of this study, we consider two different types of input spectra. All considered spectra are characterized by their spectral resolving power (\R{}) and signal-to-noise ratio (\SN{}). We define the \R{} of a spectrum as $\lambda/\Delta\lambda$, where $\Delta\lambda$ is the width of a given wavelength bin and $\lambda$ the wavelength at the bin's center. We introduce the simulated 1D Earth spectrum used to test the different forward models in Section~\ref{subsubsec:proof_concept_inputs}. Next, in Section~\ref{subsubsec:input_real_earth}, we provide information on the empirical, disk-integrated Earth spectra and the corresponding ground truths studied in the main analysis of this work. Last, we provide details on the assumptions made for the wavelength dependent \SN{} of all spectra in Section~\ref{SubSubSec:NoiseModels}.}

\subsubsection{Simulated 1D Earth spectrum}\label{subsubsec:proof_concept_inputs}

\begin{figure*}
   \centering
    \includegraphics[width=\textwidth]{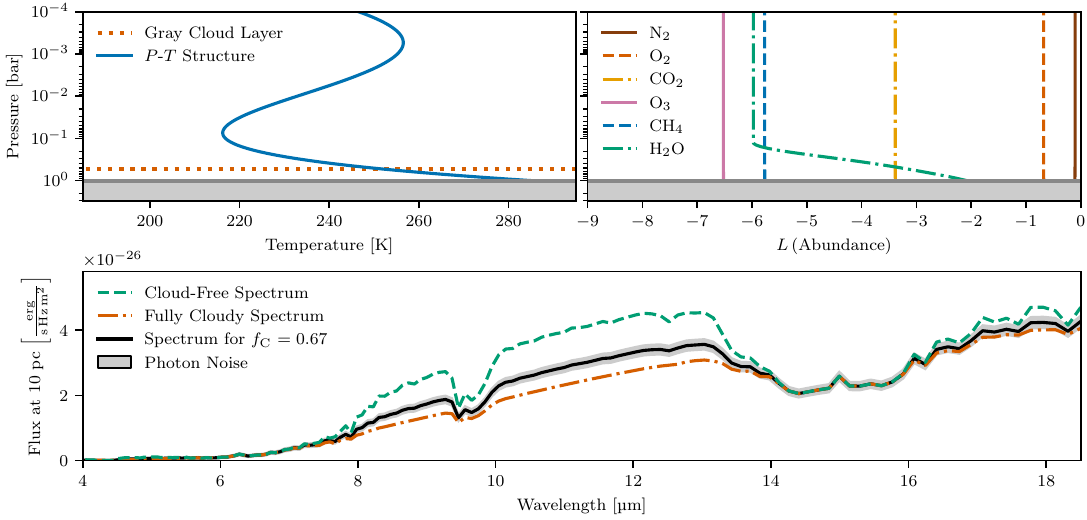}
    \caption{\pt{} and abundance profiles assumed for the simulated Earth input spectrum (Section~\ref{subsubsec:proof_concept_inputs}). \textit{Top-left panel:} Assumed Earth-like \pt{} structure (solid blue line) and position of the cloud layer (dotted red line). \textit{Top-right panel:} Vertical profiles of the considered atmospheric gases. \textit{Bottom panel:} Calculated cloud-free (dashed green line), cloudy (dashed-dotted red line), and partially cloudy ($\fcloud{}=0.67$; solid black line) Earth spectra. The gray-shaded area indicates the \SNv{20} photon noise level assumed for the partially cloudy input spectrum.}
    \label{fig:proof_concept_input}
\end{figure*}

    Before performing retrievals on real, disk-integrated Earth spectra, we run test retrievals with all forward models on a simulated spectrum of a 1D Earth-like atmosphere to validate our approach. We generate the 1D Earth spectrum using the cloudy \vwcl{} forward model (see Section~\ref{subsec:atmospheric_models}). We assume an Earth-like \pt{} structure, radius, mass, and atmospheric composition (see Table~\ref{tab:model_parameters} and Figure~\ref{fig:proof_concept_input}). The atmospheric \ce{H2O} content is determined with the condensation model by assuming $\drying{}=10^{-2}$. {For the cloud fraction, we assume $\fcloud=0.67$, which corresponds to the lower limit for the mean annual cloud fractions of the Earth views in \citet{Mettler2024} (range: $\fcloud=0.67-0.71$). We calculate the spectrum for the minimal \life{} wavelength range (\mic{4-18.5}) specified in \citet{LIFE_III}, choose a spectral resolving power of \Rv{100}, and assume \SNv{20} photon-noise (low-noise model in Section~\ref{SubSubSec:NoiseModels}). We show the obtained simulated MIR Earth spectrum in the bottom panel of Figure~\ref{fig:proof_concept_input}.}

\subsubsection{Real disk-integrated Earth spectra and ground truths}\label{subsubsec:input_real_earth}

\begin{figure*}
   \centering
    \includegraphics[width=\textwidth]{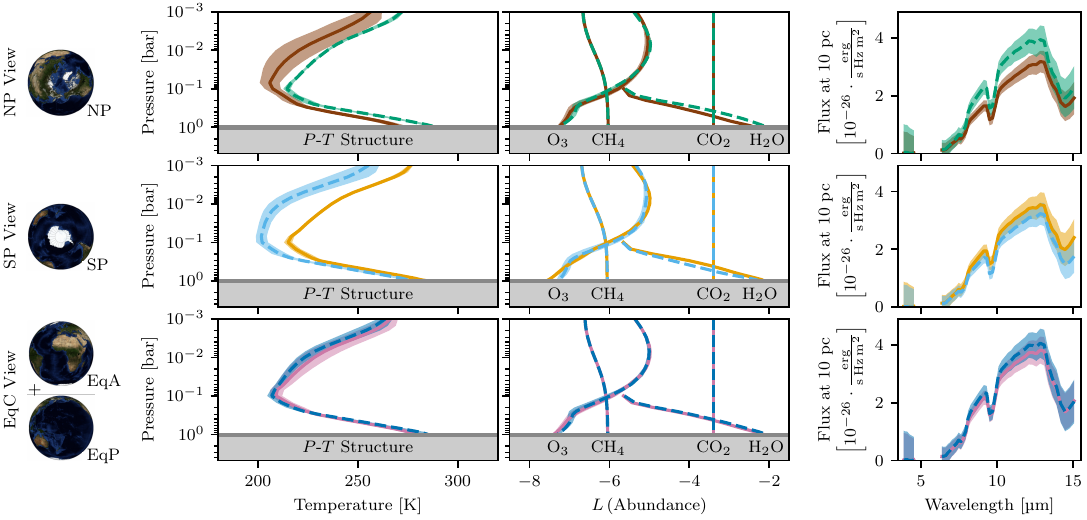}
    \caption{Ground truths and disk-integrated MIR spectra for the NP, SP, and EqC Earth views introduced in Section~\ref{subsubsec:input_real_earth}. Solid lines indicate the data for January and dashed lines represent the month July. Color-shaded areas indicate the uncertainties on the data. \textit{Left column:} Ground truth \pt{} profiles. \textit{Center column:} Ground truth abundance profiles for \ce{CO2},\ce{O3},\ce{CH4}, and \ce{H2O}. \textit{Right column:} \Rv{50} disk-integrated monthly averaged Earth spectra. The color-shaded areas indicate the \SNv{10} \lifesim{}-noise level.}
    \label{fig:disk_integrated_input}
\end{figure*}

    As input for the main retrieval analysis of this study, we use the same set of monthly averaged\footnote{We consider monthly averages since this time span roughly corresponds to the expected \life{} integration times specified in \citet{LIFE_III}.} disk-integrated Earth spectra as in \citet{Mettler2024}. The spectra are derived from Earth remote sensing climate data from NASA's Atmospheric Infrared Sounder \citep[AIRS;][]{AIRS} aboard the Aqua satellite. We refer to \citet{mettler_2023} and \citet{Mettler2024} for a description of the methods used to derive these disk-integrated, monthly averaged Earth spectra.
    
    The spectra cover the \mic{3.75-15.4} wavelength range, with a gap at \mic{4.6-6.2} due to dead instrument channels. We consider three different orientations of Earth relative to the observer. For the North (NP) and South Pole (SP) case, the respective polar region is centered on the observed disk. The Equatorial Combined (EqC) view is centered on Earth's equator and is the combination of an Africa-centered (EqA) and a Pacific-centered (EqP) equatorial view. We combine the EqA and EqP views into the EqC spectrum to account for Earth's rotation, which is significantly faster than the assumed 1-month integration time. To capture Earth's largest variability, we consider the monthly averages for January (Jan) and July (Jul) from 2017 for each orientation. This yields a total of six different Earth spectra.

    {Motivated by first \life{} requirement estimates from \citet{LIFE_III} (\Rv{50}, \SNv{10}), we consider two resolving powers (\Rv{50, 100}) and two noise levels (\SNv{10, 20}) for each spectrum. With the \life{} noise model from Section~\ref{SubSubSec:NoiseModels}, we determine the wavelength-dependent \SN{} of the disk-integrated spectra. In Figure~\ref{fig:disk_integrated_input} we show all six disk-integrated Earth Spectra for the \Rv{50}, \SNv{10} case.}

    In Section~\ref{sec:results}, we compare the retrieved posteriors with {ground truth averages} for the different views. The ground truths for the \pt{} profiles and the trace gas abundances of \ce{H2O}, \ce{CH4}, and \ce{O3} were derived from the Aqua/AIRS L3 Monthly Standard Physical Retrieval (AIRS-only) 1 degree x 1 degree V7.0 product \citep{AIRS3STM_v07}. {Ground truth averages} for \ce{CO2} were calculated from the OCO-2 GEOS Level 3 monthly dataset \citep[0.5x0.625 assimilated \ce{CO2} V10r at GES DISC;][]{OCO-2_v10}, which is derived from observations from the Orbiting Carbon Observatory 2 (OCO-2). We display the ground truths for all views and months in Figure~\ref{fig:disk_integrated_input}. For details on the calculation of these disk-integrated ground truths, we refer to \citet{Mettler2024}.

\subsubsection{Considered noise models}\label{SubSubSec:NoiseModels}

\begin{figure}
  \centering
    \includegraphics[width=0.47\textwidth]{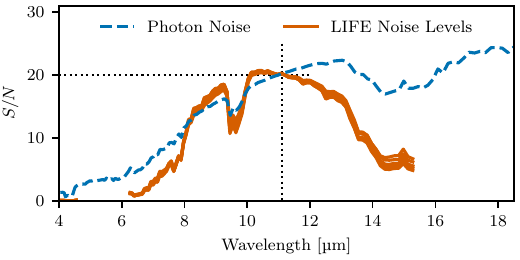}
  \caption{{Noise models for the \Rv{100}, \SNv{20} spectra. The blue-dashed line shows the photon noise assumed for the simulated Earth spectrum. The \lifesim{} noise assumed for the disk-integrated Earth spectra is shown as solid-red lines. Both noise models reach an \SN{} of 20 at \mic{11.2} (indicated by the black-dotted lines).}}
  \label{fig:noise_models}
\end{figure}

    {In the present study, we use two different approaches to model the wavelength-dependent \SN{} of the considered MIR spectra. First, to test the different forward models on the simulated Earth spectrum, we assume a low-noise scenario. Second, for the retrievals on the disk-integrated Earth spectra, we use a more realistic \life{}-like \SN{} model. Because the \SN{} is not constant over the spectrum for both cases, we define the \SN{} of the spectrum as the \SN{} in the \mic{11.2} wavelength bin (see Figure~\ref{fig:noise_models}). We choose the \mic{11.2} bin since it does not coincide with strong absorption features and thus probes Earth's continuum emission. This reference wavelength was also used by previous \life{}-related retrieval studies \citep{LIFE_III, LIFE_V, LIFE_IX, Mettler2024, Alei2024arX}.}

    {For the low-noise scenario, we only assume photon noise of the source to be present. Thus, if $N$ photons are detected in a given wavelength bin, we assume the noise to be $\sqrt{N}$. To calculate the \SN{}, we scale the flux such that the desired \SN{} is reached at \mic{11.2}. The wavelength-dependent photon noise is indicated by the blue-dashed line in Figure~\ref{fig:noise_models}.}

    {To obtain the more realistic \life{}-like noise estimates, we use the \lifesim{} model \citep{LIFE_II}. In addition to accounting for photon noise of the planet emission, \lifesim{} models contributions from major astrophysical noise sources (stellar leakage and local- as well as exozodiacal dust emission). Thus, \lifesim{} makes the implicit assumption that observations with a \life{}-like instrument will not be dominated by instrumental noise terms (Dannert et al., in prep.; Huber et al., in prep.). To calculate the \lifesim{} noise, we assume the observed planet to orbit a G2V star at $1$~AU at a separation of $10$~pc from us. The exozodiacal dust emission is set to three times the local zodiacal dust emission, which is the median exozodi emission found for Sun-like stars by the HOSTS survey \citep{ertel2020}. The wavelength-dependence of the \lifesim{} noise we assume for the disk-integrated Earth spectra is indicated by the solid-red lines in Figure~\ref{fig:noise_models}.}

    {For all spectra, we treat the wavelength-dependent \SN{} as uncertainty on the flux. This implies that the spectral points are set to the true flux values and are not randomized according to the \SN{}. While randomized spectra provide more accurate simulated observations, retrievals on such noise realizations yield biased parameter posteriors. Ideally, our retrieval study would consider a large number ($\gtrsim10$) of different noise realizations of one spectrum. However, such a study is computationally unfeasible due to the large number of required atmospheric retrievals. Alternatively, as motivated in \citet{LIFE_III}, retrievals on unrandomized spectra can be used to obtain estimates for the average expected retrieval performance for randomized spectra.}

\subsection{Prior distributions of model parameters}\label{subsec:priors}

    All prior distributions assumed to run the retrievals are listed in Table~\ref{tab:model_parameters}. We choose the priors for the \pt{} parameters $a_i$ and the surface pressure \Ps{} such that the resulting \pt{} structures cover a wide range of profiles (from cold and thin Mars-like to hot and massive Venus-like atmospheres). For the atmospheric gases \ce{N2}, \ce{O2}, \ce{CO2}, \ce{O3}, and \ce{CH4}, we assume broad uniform priors in log-space that extend from $1$ to $10^{-10}$ in mass fraction. The lower limit of $10^{-10}$ lies significantly below the estimated minimal detectable abundances presented in \citet{LIFE_III} ($\approx10^{-7}$ in mass fraction).

    Based on the retrieved abundance constraints form \citet{Mettler2024}, we exclude \ce{H2O} dominated atmospheres by setting the upper limit of $10^{-1}$ for the \ce{H2O} prior. For planets on Earth-like orbits around G-stars, high \ce{H2O} abundances can occur for high partial pressures of \ce{CO2} ($p_{\ce{CO2}}\approx0.1-1$~bar) or other greenhouse gases due to the increased surface temperature \citep[e.g.,][]{Wordsworth2013}. Yet, such atmospheres lie close to the critical runaway greenhouse limit and are out of scope for this study. For the \drying{} parameter, we choose a log-uniform prior between $1$ to $10^{-5}$. This allows for atmospheric \ce{H2O} to deplete by up to five orders of magnitude in the upper atmosphere. Further, the uniform \fcloud{} prior between 0 and 1 covers the full range from cloud-free to fully cloudy atmospheres.

    {We assumed Gaussian priors for the planet radius \Rpl{} and mass \Mpl{}. The \Rpl{} prior is motivated by \citet{LIFE_II}, who find that already the detection of a planet with \life{} yields a first \Rpl{} estimate. For terrestrial planets in the HZ, they predict a radius estimate $R_\mathrm{est}$ for the true radius $R_\mathrm{true}$ of $R_\mathrm{est}/R_\mathrm{true}=0.97\pm0.18$. We derive the prior for \lgrt{\Mpl{}} from the \Rpl{} prior using the statistical mass-radius relation \texttt{Forecaster} \citep{Kipping:Forecaster}. These \Rpl{} and \Mpl{} priors are in accordance with previous \life{}-related retrieval studies \citep{LIFE_III,LIFE_IX,LIFE_V,Mettler2024}.}
\section{Retrieval Results}\label{sec:results}

In the following, we present the results from the validation retrievals on the simulated 1D Earth spectrum in Section~\ref{SubSec:ResValidation}. Thereafter, the findings from the retrievals on the six empirical disk-integrated Earth views are presented in Section~\ref{SubSec:RealSpectra}.

\subsection{Results for the simulated 1D Earth spectrum}\label{SubSec:ResValidation}

\begin{figure*}
   \centering
    \includegraphics[width=\textwidth]{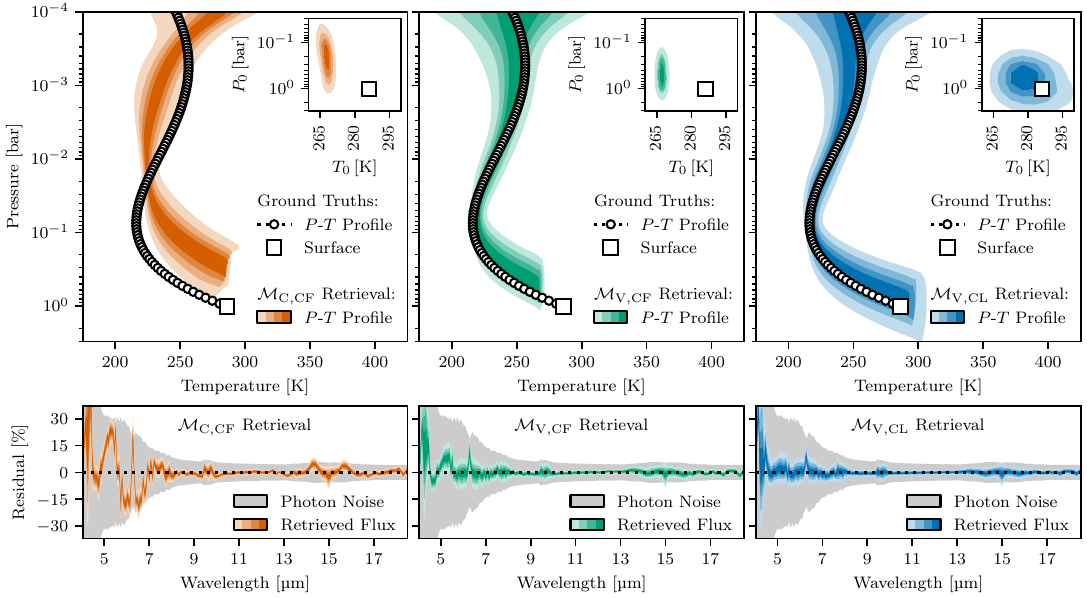}
    \caption{\pt{} fits (top row) and flux residuals (bottom row) for the retrievals on the simulated \Rv{100}, \SNv{20} Earth spectrum. Each column provides the results for a different retrieval model (from left to right: \cwcf{}, \vwcf{}, and \vwcl{}). Color-shaded areas indicate retrieval percentiles (from light to dark colors: $5-95\%$, $15-85\%$, $25-75\%$, and $35-65\%$). In the \pt{} figures, white squares mark the true surface conditions (\Ps{}, \Ts{}) and white circles the true \pt{} structure. The inlay in the top right of the \pt{} panel shows the retrieved constraints on the surface conditions. In the flux residual figures, the gray-shaded area indicates the assumed 1-$\sigma$ photon noise level.}
         \label{fig:proof_concept_PT_Residuals}
    \end{figure*}

    We present the retrieval results for the simulated 1D Earth spectrum (Section~\ref{subsubsec:proof_concept_inputs}) using the three retrieval models \cwcf{}, \vwcf{}, and \vwcl{} (Table~\ref{tab:model_summary}, Section~\ref{subsec:atmospheric_models}). In Figure~\ref{fig:proof_concept_PT_Residuals}, we illustrate the constraints retrieved for the \pt{} structure and the surface conditions. For the \cwcf{} model, the retrieved \pt{} structure is shifted to lower pressures relative to the ground truth. On average, the surface temperature \Ts{} is underestimated by $18$~K and the surface pressure \Ps{} by $0.7$~dex {(`dex' reports the value of a quantity $q$ in log-space: $\lgrt{q}=k$~dex; $q = 10^k$~Unit$(q)$)}. With the \vwcf{} model, which accounts for \ce{H2O} condensation, these shifts are significantly reduced and the ground truth lies within the 1-$\sigma$ envelope of the retrieved \pt{} structure. Yet, \Ps{} remains slightly underestimated ($\lgrt{\Ps{}}=0.3\pm0.2$~dex) and the \Ts{} constraint is not improved. The \vwcl{} model, which assumes a partial cloud coverage, yields the most accurate constraints for the \pt{} structure. The true \Ps{} lies within the 16\%-84\% percentile range of the posterior ($\lgrt{\Ps}=0.2\pm0.3$~dex). The estimate for \Ts{} is also improved significantly, but the uncertainty is increased ($\Ts{}=280\pm8$~K).

    The flux residuals in Figure~\ref{fig:proof_concept_PT_Residuals} show that the \cwcf{} model fails to accurately fit the input spectrum. Especially in the strong absorption features of \ce{H2O} (\mic{4.5-7.5}), \ce{O3} (\mic{9-10}), and \ce{CO2} (\mic{14-16}), the fitted flux differs significantly from the input, which indicates these features cannot be modeled correctly. In contrast, the \vwcf{} retrieval provides an accurate fit to the input spectrum above $\sim\mic{5.5}$, which is in agreement with the improvements observed for the \pt{} structure. Finally, the \vwcl{} model yields further minor improvements over the \vwcf{} model in the high noise regime of the input spectrum between \mic{4-6}.

\begin{figure}
   \centering
    \includegraphics[width=0.47\textwidth]{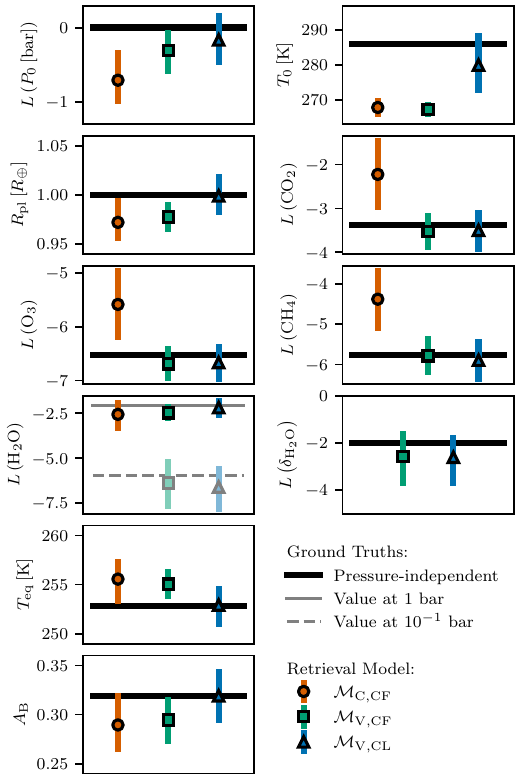}
    \caption{Posteriors from retrievals on the simulated \Rv{100}, \SNv{20} Earth spectrum. Here, $L(\cdot)$ stands for $\lgrt{\cdot}$. Thick black lines indicate pressure-independent ground truths. Thin gray lines show the true \ce{H2O} abundance at 1~bar (solid line) and at $10^{-1}$~bar (dashed line). The different markers represent the retrieval models: orange circles -- \cwcf{}, green squares -- \vwcf{}, and blue triangles -- \vwcl{}. The colored lines indicate the $16\%-84\%$ posterior percentiles. For the \vwcf{} and \vwcl{} models, which assume variable \ce{H2O} profiles, we provide both the retrieved \ce{H2O} abundance at \Ps{} (dark marker color) and at 1~dex below \Ps{} (light marker color).}
         \label{fig:proof_concept_posteriors}
    \end{figure}

    In Figure~\ref{fig:proof_concept_posteriors} and Table~\ref{Table:Proof_of_concept posteriors}, we summarize the retrieved posteriors for the forward model parameters. We provide the posteriors for the surface conditions, the planet radius \Rpl{}, the atmospheric trace gases, the drying parameter \drying{}, and the cloud fraction \fcloud{}. The $a_i$ posteriors are visualized by the retrieved \pt{} structures in Figure~\ref{fig:proof_concept_PT_Residuals}. The posteriors for \Mpl{}, \ce{N2}, and \ce{O2} are not shown, since they were not constrained over the assumed priors. {Last, we derived estimates for the planet's equilibrium temperature \Teq{} and Bond albedo \Ab{} from the posteriors. For \Teq{}, we calculated the MIR spectra corresponding to the points in the posterior and integrated these spectra to obtain the total MIR emission. We then defined \Teq{} as the temperature of a spherical black-body with radius \Rpl{} that emits the same total flux. From \Teq{}, we then determined \Ab{}:}
    \begin{equation}\label{eq:bond_albedo}
        \Ab = 1 - 16\,\pi\sigma \frac{a^2_P\Teq^4}{L_*}.
    \end{equation}
    {Here, $a_P$ is the exoplanet's semi-major axis and $L_*$ the star's luminosity. For details on the derivation of \Teq{} and \Ab{}, we refer to \citet{Mettler2024}.}

    \begin{deluxetable}{ccccc}
\tablecaption{Numeric values for the posteriors in Figure~\ref{fig:proof_concept_posteriors}.}
\label{Table:Proof_of_concept posteriors}      
\tablehead{
\colhead{\multirow{2}{*}{Parameter}}&\colhead{Ground}&\multicolumn{3}{c}{Posteriors}\\\cline{3-5}
&Truths&\cwcf{}&\vwcf{}&\vwcl{}
}
\startdata
    $L\left(P_0\left[\mathrm{bar}\right]\right)$ &{$0.0$} &$-0.7^{+0.4}_{-0.3}$ &$-0.3^{+0.2}_{-0.3}$ &$-0.2^{+0.3}_{-0.3}$ \\
    $T_0\left[\mathrm{K}\right]$ &{$286$} &$268^{+2}_{-2}$ &$267^{+1}_{-1}$ &$280^{+8}_{-7}$ \\
    $R_\mathrm{pl}\left[R_\oplus\right]$ &{$1.00$} &$0.97^{+0.02}_{-0.02}$ &$0.98^{+0.01}_{-0.01}$ &$1.00^{+0.02}_{-0.02}$ \\
    $L\left(\mathrm{CO_2}\right)$ &{$-3.4$} &$-2.2^{+0.8}_{-0.7}$ &$-3.5^{+0.4}_{-0.3}$ &$-3.5^{+0.4}_{-0.4}$ \\
    $L\left(\mathrm{O_3}\right)$ &{$-6.5$} &$-5.6^{+0.6}_{-0.6}$ &$-6.7^{+0.3}_{-0.3}$ &$-6.7^{+0.3}_{-0.3}$ \\
    $L\left(\mathrm{CH_4}\right)$ &{$-5.8$} &$-4.4^{+0.7}_{-0.7}$ &$-5.8^{+0.4}_{-0.4}$ &$-5.9^{+0.5}_{-0.5}$ \\
    $L\left(\mathrm{H_2O}\right)$ &$-2.1$ &$-2.6^{+0.7}_{-0.8}$ &$-2.5^{+0.3}_{-0.3}$ &$-2.2^{+0.4}_{-0.4}$ \\
    $L\left(\delta_\mathrm{H_2O}\right)$ &{$-2.0$} &... &$-2.6^{+0.9}_{-1.1}$ &$-2.6^{+0.8}_{-1.1}$ \\
    $f_\mathrm{C}$ &{$0.67$} &... &... &$0.54^{+0.17}_{-0.21}$\\
    $T_\mathrm{eq}\left[\mathrm{K}\right]$ &{$253$} &$256^{+2}_{-2}$ &$255^{+1}_{-1}$ &$253^{+2}_{-2}$ \\
    $A_\mathrm{B}$ &$0.32$ &$0.29^{+0.03}_{-0.02}$ &$0.29^{+0.02}_{-0.02}$ &$0.32^{+0.02}_{-0.02}$
\enddata
\tablecomments{Here, $L(\cdot)$ stands for $\lgrt{\cdot}$. The second column lists the assumed ground truths (see Table~\ref{tab:model_parameters} for details). In the last three columns, we list the median and the $16\% - 84\%$ range (via $+/-$ indices) of the parameter posteriors for the \cwcf{}, \vwcf{}, and \vwcl{} models. For \ce{H2O}, all listed values correspond to the surface-layer abundances.}
\end{deluxetable}
    
    As seen for the \pt{} structures, both the \Ps{} and \Ts{} posteriors are biased relative to the ground truth for the \cwcf{} retrieval. Due to a degeneracy between the pressure-induced line-broadening and the abundances of the atmospheric trace-gases \citep[see, e.g.,][]{Misra2014,Schwieterman2015,Alei2022}, we expect the underestimation of the pressures to be accompanied by an overestimation of the trace-gas abundances. Indeed, the \ce{CO2}, \ce{O3}, and \ce{CH4} abundances are overestimated by $0.9$~dex to $1.4$~dex by the \cwcf{} retrieval. For the \vwcf{} and \vwcl{} retrievals, which yield better pressure constraints, the abundance estimates are unbiased (i.e., the ground truths fall within the 16\%-84\% posterior range). Only the \ce{H2O} abundance, which decreases strongly with altitude in Earth's atmosphere, is not overestimated by the \cwcf{} retrieval. The retrieved abundance lies between the 1~bar and $10^{-1}$~bar ground truth, roughly $0.5$~dex below the surface abundance (see Table~\ref{Table:Proof_of_concept posteriors}). For the \vwcf{} retrieval, the abundance at $10^{-1}$~bar is accurately estimated. Yet, since \Ps{} and \Ts{} are underestimated, the condensation model underestimates the surface \ce{H2O} abundance by $0.4$~dex. With the \vwcl{} model, this bias on the surface abundance of \ce{H2O} is removed. The \drying{} estimates from the \vwcf{} and \vwcl{} scenarios do not differ significantly. In both cases, values of $\drying{}>10^{-1}$ are ruled out. Thus, an Earth-like \ce{H2O} depletion in the upper atmosphere is in principle detectable from the MIR thermal emission.

    Similar to \Ts{}, \Rpl{} is underestimated by $\sim0.03~R_\oplus$ by the \cwcf{} and \vwcf{} retrievals. This bias is propagated to the \Teq{} and \Ab{} estimates, since they are derived from the \Rpl{} posterior. For the \cwcf{} and \vwcf{} retrievals, the underestimated \Rpl{} translates to an overestimated \Teq{} (by $\sim3$~K) and an underestimated \Ab{} (by $\sim0.03$). Thus, a too small \Rpl{} is compensated with an increased \Teq{}, which requires a lower \Ab{}. In contrast, the retrievals using the true \vwcl{} model yield accurate and bias-free estimates for \Rpl{}, \Teq{}, and \Ab{}.

\begin{figure}
   \centering
    \includegraphics[width=0.465\textwidth]{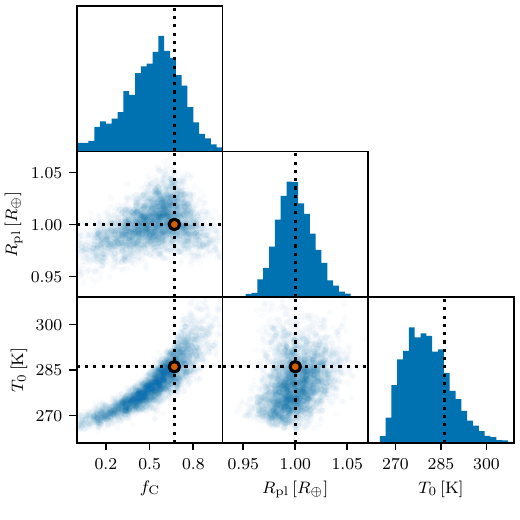}
    \caption{Corner plot for the \fcloud{}, \Rpl{}, and \Ts{} posteriors (in blue) for the retrieval with the \vwcl{} forward model on the simulated \Rv{100}, \SNv{20} Earth spectrum. The black-dashed lines and the circular orange markers indicate the parameter ground truths.}
        \label{fig:proof_concept_correlations}
    \end{figure}

    Last, in the \vwcl{} retrievals, we aim to constrain the cloud fraction \fcloud{}. We show the \fcloud{}, \Rpl{}, and \Ts{} posteriors from the \vwcl{} retrieval in the corner plot in Figure~\ref{fig:proof_concept_correlations}. While \fcloud{} is not strongly constrained over the prior range, 
    we observe significant correlations between \fcloud{} and the other parameters. The correlation between \fcloud{} and \Ts{} implies that a higher \Ts{} can be compensated with a higher \fcloud{}, since atmospheric clouds block the thermal emission from the warm near-surface layers. This degeneracy explains the observed increase in the \Ts{} uncertainty for the \vwcl{} retrieval. Further, \fcloud{} is weakly correlated with \Rpl{} (higher \fcloud{} lead to higher \Rpl{}), which explains why \Rpl{} estimates are improved for the \vwcl{} model. 

    Finally, we use Bayes' factor $K$ (values in Table~\ref{Table:Evidences_and_K}) to benchmark the models against each other. We find that \vwcf{} is preferred over \cwcf{} ($|\lgrt{K}|=3.7\pm0.1$), and \vwcl{} outperforms \cwcf{} ($|\lgrt{K}|=4.3\pm0.1$). Thus, an accurate model for Earth's \ce{H2O} abundance profile is required to correctly fit its MIR spectrum. This agrees with our findings for the retrieved \pt{} structures, flux residuals, and parameter posteriors. Further, \vwcl{} is slightly preferred over the cloud-free \vwcf{} model ($|\lgrt{K}|=0.6\pm0.1$). This slight preference agrees with the small improvements observed for the flux residuals. Yet, despite the limited improvements in the spectral fit, the degeneracies associated with \fcloud{} help improve the estimates for \Ts{}, \Rpl{}, \Teq{}, and \Ab{}.

\subsection{Results for real disk-integrated Earth spectra}\label{SubSec:RealSpectra}

\begin{figure*}
   \centering
    \includegraphics[width=\textwidth]{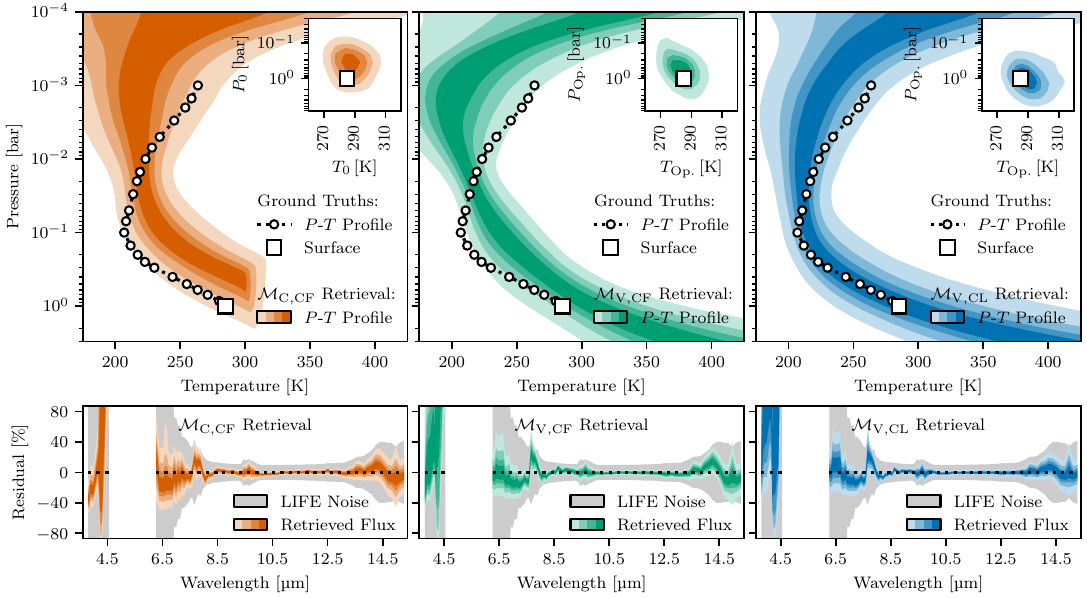}
    \caption{\pt{} fits (top row) and flux residuals (bottom row) for the retrievals on the \Rv{100}, \SNv{10} EqC~Jul spectrum of Earth. Each column provides the results for a different retrieval model (from left to right: \cwcf{}, \vwcf{}, and \vwcl{}). Color-shaded areas indicate retrieval percentiles (from light to dark colors: $5-95\%$, $15-85\%$, $25-75\%$, and $35-65\%$). In the \pt{} figures, white squares mark the true surface conditions (\Ps{}, \Ts{}) and white circles the true \pt{} structure. The inlay in the top right of the \pt{} panel shows the retrieved constraints on the surface conditions (\Ps{}, \Ts{}) for the \cwcf{} case. For the \vwcf{} and \vwcl{} scenario, the inlay shows the pressure and temperature where the atmosphere becomes opaque (\Popaque{}, \Topaque{}). In the flux residual figures, the gray-shaded area indicates the assumed 1-$\sigma$ \lifesim{}-noise level.}
         \label{fig:View_examle_PT}
    \end{figure*}

    Here, we present the retrieval results for the empirical disk-integrated Earth views (Section~\ref{subsubsec:input_real_earth}) for the \cwcf{}, \vwcf{}, and \vwcl{} retrieval models (Table~\ref{tab:model_summary}, Section~\ref{subsec:atmospheric_models}). In Figure~\ref{fig:View_examle_PT}, we show the retrieved \pt{} structures for the \Rv{100}, \SNv{10} EqC~Jul Earth spectrum. Further, we show the residuals of the fitted spectra relative to the input spectrum. The \pt{} results and the residuals for {all other views}, \R{}, and \SN{} scenarios are comparable and thus not shown.

    The \pt{} structures retrieved for the \cwcf{} model exhibit similar characteristics as the results for the simulated spectrum. The retrieved constraints are systematically shifted toward lower pressures. Also \Ps{} is underestimated by $0.4$~dex to $0.6$~dex depending on the view. In contrast, \Ts{} is well approximated by the \cwcf{} retrievals (truths lie in the $16\%-84\%$ posterior range). While shifts in the retrieved \pt{} profiles relative to the ground truth are also observed for \vwcf{}, the \vwcl{} runs yield accurate constraints. However, for both \vwcf{} and \vwcl{}, the \pt{} constraints, which extend significantly beyond the true \Ps{} and \Ts{}, indicate that the surface conditions are no longer well constrained. {This suggests that the high-pressure layers of the fitted model atmosphere do not contribute significantly to the MIR thermal emission emerging at the top of the atmosphere. Alternatively, we can consider the pressure (\Popaque{}) and temperature (\Topaque{}) in the deepest layer in the model atmosphere that still contributes to the simulated MIR spectrum. Thus, \Popaque{} and \Topaque{} correspond to the pressure and temperature at which the simulated atmosphere becomes opaque in the MIR. For an optically thin atmosphere, we simply obtain: $\Popaque{}=\Ps{}$ and $\Topaque{}=\Ts{}$. We observe that \Popaque{} and \Topaque{} roughly correspond to the ground truths for \Ps{} and \Ts{}. Thus, the fitted model atmospheres become opaque close to Earth's true surface.} 
    

    The flux residuals in Figure~\ref{fig:View_examle_PT} do not differ significantly between the models. Around the \mic{7.7} \ce{N2O} band, the models struggle to fit the spectrum accurately, since they do not include \ce{N2O}. \citet{Mettler2024} showed that, for the empirical disk-integrated Earth spectra we consider here, models including \ce{N2O} are not preferred by Bayesian model selection. At all other wavelengths, the three models provide an accurate fit. This observation is supported by model comparison via the Bayes factor $K$ (values in Table~\ref{Table:Evidences_and_K}). For all \Rv{50} and most \Rv{100} spectra, no model is preferred over the others ($|\lgrt{K}|\leq0.3$; uncertainty $\pm0.2$). Rare exceptions occur for the \Rv{100}, \SNv{20} spectra, where slight preferences for \vwcf{} or \vwcl{} are occasionally observed ($|\lgrt{K}|\leq0.6$; uncertainty $\pm0.2$).

\begin{figure*}
   \centering
    \includegraphics[width=\textwidth]{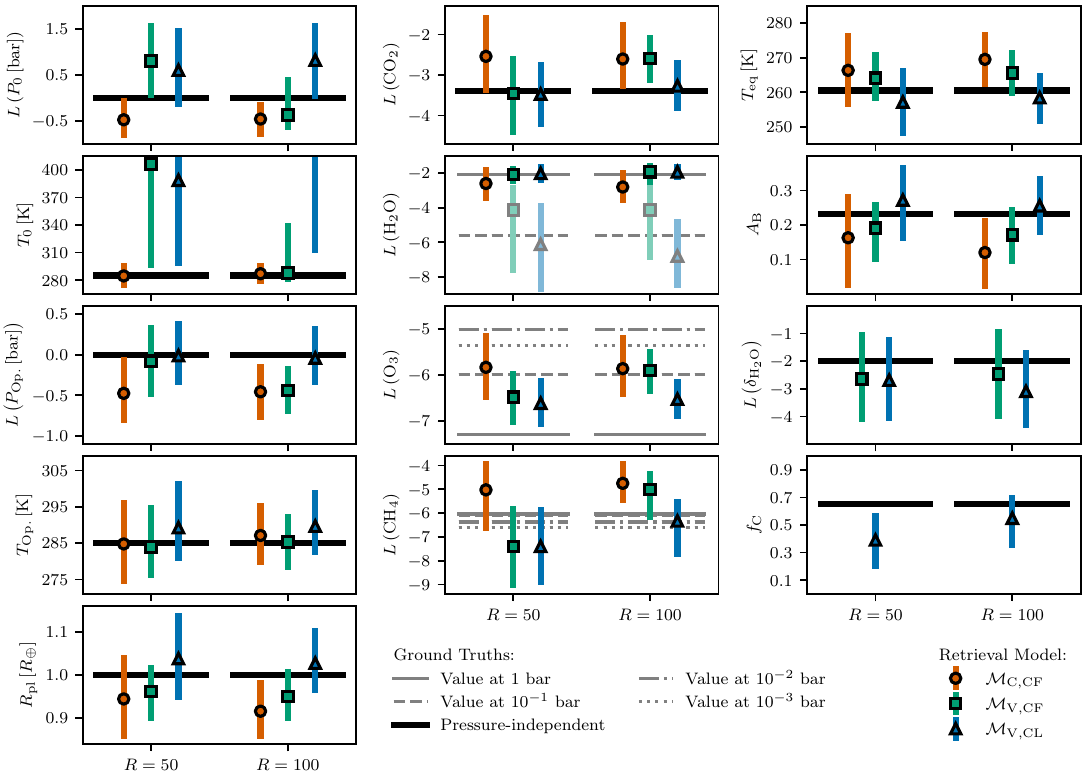}
    \caption{{Posteriors from the retrievals on the EqC~Jul spectra of Earth (\lifesim{} \SNv{10}; \Rv{50,100}). Here, $L(\cdot)$ stands for $\lgrt{\cdot}$. Thick black lines indicate pressure-independent ground truths. Thin gray lines show the ground truth abundances at different atmospheric pressures (solid: 1~bar; dashed: $10^{-1}$~bar; dashed-dotted: $10^{-2}$~bar; dotted: $10^{-3}$~bar). The markers represent the retrieval models: orange circles -- \cwcf{}, green squares -- \vwcf{}, and blue triangles -- \vwcl{}. The colored lines indicate the $16\%-84\%$ posterior percentiles. For the \vwcf{} and \vwcl{} models, which assume variable \ce{H2O} profiles, we provide both the retrieved \ce{H2O} abundance at the pressure \Popaque{} where the atmosphere becomes optically thick (dark marker color) and at 1~dex below \Popaque{} (light marker color).}}
         \label{fig:EqCJul_posteriors_SN10}
    \end{figure*}

    {In Figure~\ref{fig:EqCJul_posteriors_SN10}, we visualize the posteriors retrieved with the \cwcf{}, \vwcf{}, and \vwcl{} models for the EqC~Jul view at a \lifesim{} \SN{} of $10$. The results for the other views and the \SNv{20} retrievals exhibit the same general behavior and are provided in Figures~\ref{fig:views_posteriors_SN10} and \ref{fig:views_posteriors_SN20} in Appendix~\ref{App:additional_results}. Further, the numeric values for all posteriors are listed in Tables~\ref{tab:Ret_NP}, \ref{tab:Ret_SP}, and \ref{tab:Ret_EqC}. We show the posteriors for \Ps{}, \Ts{}, \Popaque{}, \Topaque{}, \Rpl{}, the trace-gas abundances, \drying{}, and \fcloud{}. The $a_i$ posteriors are visualized via the \pt{} constraints in Figure~\ref{fig:View_examle_PT}. We do not show the \Mpl{}, \ce{N2}, and \ce{O2} posteriors, since they are not constrained over the assumed priors. The \Teq{} and \Ab{} estimates were derived from the posteriors (see Section~\ref{SubSec:ResValidation}).}
    
    {We observe that the posteriors for the different models exhibit trends similar to the ones observed for the simulated 1D Earth spectra in Section~\ref{SubSec:ResValidation}.} For the \cwcf{} model, the underestimated \Ps{} values are accompanied by overestimated abundances of \ce{CO2} (by $\geq0.6$~dex) and \ce{CH4} (by $\geq1.0$~dex). For \ce{O3}, the pressure-dependent ground truths cover a wide range of abundances. Thus, the pressure-induced biases are not directly observable. Next, the \cwcf{} posteriors for \ce{H2O} underestimate the true surface-layer abundances by $0.5$~dex to $1.0$~dex. Further, \Rpl{} is consistently underestimated by $\geq0.05~R_\oplus$, which translates to biases for the derived parameters \Teq{} and \Ab{} (\Teq{} overestimated by $\geq5$~K; \Ab{} underestimated by $\geq0.05$). Only \Ts{} is accurately estimated by \Topaque{} for most scenarios.

    For \vwcf{} and \vwcl{}, significant improvements in the parameter estimates are observable. Generally, the \Ps{} estimates are higher than for the \cwcf{} runs. As seen with the \pt{} profiles, several \vwcf{} and \vwcl{} retrievals only yield lower limits for \Ps{} and \Ts{}. {In these cases, the \Popaque{} value indicates that the high-pressure layers of the model atmosphere do not contribute to the MIR spectrum and thus \Ps{} and \Ts{} cannot be constrained further}. The retrieved \Popaque{} estimates increase with the model complexity and mostly coincide with the \Ps{} ground truth for the \vwcl{} retrievals. Further, \Topaque{} provides accurate estimates for \Ts{} in the \vwcf{} retrievals. For \vwcl{}, \Topaque{} generally overestimates \Ts{} and the uncertainties are often significantly increased. This is due to the aforementioned degeneracy between \fcloud{} and \Ts{}, where too high \Ts{} can be compensated by a higher \fcloud{}. Also, the retrieved estimates for \Rpl{} and the derived parameters \Teq{} and \Ab{} are significantly improved, especially for the \vwcl{} retrievals.
    
    For the trace-gas abundances, the biases observed for the \cwcf{} retrievals are continually reduced as we include \ce{H2O} condensation and clouds. The \vwcl{} model yields the best estimates for the atmospheric \ce{CO2} abundances. For all \vwcl{} cases considered, the \ce{CO2} ground truths lie in the $16\%-84\%$ percentile range of the posteriors. Also for \ce{CH4}, the biases are reduced and the \vwcl{} posteriors estimate the ground truths accurately for most scenarios. Yet, for some \Rv{50}, \SNv{10} scenarios, \ce{CH4} is no longer detectable. In these cases, only an upper limit is obtained for the \ce{CH4} abundance ($\approx$$-6.0\pm0.2$~dex). For \ce{O3}, we observe a significant reduction in the retrieved abundance. This indicates that the \ce{O3} estimate in the \cwcf{} retrieval is affected by the same pressure-abundance degeneracy as \ce{CO2} and \ce{CH4}. Last, the \ce{H2O} abundance at \Popaque{} provides accurate estimates for Earth's ground truth surface abundance of \ce{H2O}. Further, the structure of the \ce{H2O} profile is accurately approximated by both the \vwcf{} and \vwcl{} retrievals.  
    
    Lastly, all retrievals yield similar constraints for \drying{} and \fcloud{} independent of the disk-integrated Earth spectrum considered. As for the simulated Earth spectrum in Section~\ref{SubSec:ResValidation}, only high values of $\drying{}>10^{-1}$ can be ruled out for both the \vwcf{} and the \vwcl{} model. {Further, the \vwcl{} retrievals often struggle to constrain \fcloud{} due to the aforementioned correlations with \Topaque{}. Generally, values of $\fcloud{}\approx0.5\pm0.2$ are retrieved and only $\fcloud{}$ lower than $0.1$ can be ruled out confidently. In some \SNv{20} retrievals, the lower limit obtained for $\fcloud{}$ is significantly higer (lower limit on $\fcloud{}$: $\approx0.5$).}
\section{Discussion}\label{sec:discussion}

    {Here, we outline implications of our work for the characterization of terrestrial HZ exoplanets (Section~\ref{SubSec:implications_characterization}) and for the \life{} mission requirements (Section~\ref{SubSec:Implications_LIFE}). Thereafter, in Section~\ref{SubSec:limitations_future_work}, we point out the limitations of the present study.}

\subsection{Implications for Characterizing Rocky HZ Exoplanets}\label{SubSec:implications_characterization}

    {We find that the biased estimates for the \pt{} structure and the atmospheric composition obtained in the \cwcf{} retrievals are predominantly attributable to two simplifying assumptions commonly made by retrieval models: vertically constant abundance profiles and a cloud-free atmosphere. Our results for the \vwcf{} and \vwcl{} retrieval models demonstrate that including physically motivated models for Earth's atmospheric \ce{H2O} profile and partial cloud coverage helps significantly improve the accuracy of the retrieved parameter estimates. However, despite the improved parameter estimates, neither the \vwcf{} nor the \vwcl{} model is preferred over the \cwcf{} model by Bayesian model comparison for \life{}-like \R{} and \SN{} scenarios. Yet, we argue that the usage of the \vwcf{} and \vwcl{} models is justifiable, since both impose physics informed constraints on the model atmosphere. This allows to exclude unrealistic atmospheric states and thereby yields more reliable parameter estimates. Especially for HZ exoplanets, unbiased parameter constraints are a critical requirement to enable the accurate assessment of their habitability. Further, in the same context, biased estimates for the abundances of potential biosignature gases, such as \ce{CH4}, can be highly problematic.}

    {From Figure~\ref{fig:disk_integrated_input} and \citet{mettler_2023}, we know that Earth's MIR emission spectrum depends significantly on both the observed view and the season. Since both the viewing geometry and the season will be unknown for future observations of HZ terrestrial exoplanets, it is crucial to understand if and how these factors affect our characterization. We find that, independent of the assumed retrieval model, the retrieved constraints for most of the model parameters exhibit no significant dependence on the view and season of the Earth spectrum. Only \Ts{}, \Topaque{}, \Teq{}, and \Ab{} exhibit a perceptible dependence on the considered spectrum in the high \R{} and \SN{} scenarios (see Figures~\ref{fig:views_posteriors_SN10} and \ref{fig:views_posteriors_SN20}). These findings are in agreement with \citet{Mettler2024}, who performed retrievals on the same set of disk-integrated Earth spectra considered here using the \cwcf{} forward model.}

\subsection{Implications for the LIFE Initiative}\label{SubSec:Implications_LIFE}

    Several retrieval studies have evaluated how well rocky HZ exoplanets could be characterized from their MIR spectrum by a \life{}-like observatory. In \citet{LIFE_III}, preliminary minimal requirements for \life{} are obtained by analyzing simulated Earth spectra. By requiring a detection of the potential biosignature \ce{CH4} with \life{}, the authors find lower limits of \mic{4-18.5} for the wavelength coverage, $50$ for \R{}, and $10$ for \SN{}. In two follow-up studies \citep{LIFE_V,LIFE_IX}, these findings are reevaluated on simulated spectra of Earth at different temporal epochs from \citet{Rugheimer_2018} and a cloudy Venus-like planet. Last, \citet{Mettler2024} test the \life{} requirements on the same real disk- and time-averaged MIR Earth spectra we considered here.

    While the analyzed spectra vary significantly in complexity, all previous studies assumed retrieval models with vertically constant abundance profiles. Further, only \citet{LIFE_IX} accounted for clouds in their retrieval models. All studies yield similar results for the strengths of the parameter constraints and the detectability of trace-gases. They find Earth-like levels of \ce{H2O}, \ce{CO2}, \ce{O3}, and \ce{CH4} to be detectable. Further, the retrieved $16\%-84\%$ posterior range does not exceed $\pm0.5$~dex for pressures, $\pm20$~K for temperatures, $\pm0.1\,R_\oplus$ for \Rpl{}, and $\pm1.0$~dex for trace-gas abundances. These constraints from the previous \life{} studies are consistent with our results for the \cwcf{} retrieval model, which also assumes constant abundance profiles and neglects clouds.

    However, similar to our findings for \cwcf{} retrievals, both \citet{LIFE_V} and \citet{Mettler2024} observe significant biases in the retrieved posteriors relative to the ground truths. Both studies analyzed spectra of cloudy atmospheres with non-constant trace-gas abundances. \citet{Mettler2024} argue that underestimated pressures and overestimated trace-gas abundances are primarily evoked by the retrieval model assumption of a vertically constant \ce{H2O} profile. In contrast, biased temperature, \Rpl{}, and \Ab{} estimates are attributable primarily to neglecting clouds. Our findings for the simulated Earth spectrum (Section~\ref{SubSec:ResValidation}) validate these claims. In the \vwcf{} retrievals, biases on pressure and abundance estimates are reduced significantly by modeling Earth's \ce{H2O} abundance profile. Yet, temperature, \Rpl{}, and \Ab{} estimates are largely unaffected. These biases are only reduced in the cloudy \vwcl{} retrievals. Also our findings for the real disk-integrated Earth views (Section~\ref{SubSec:RealSpectra}) validate the claims from \citet{Mettler2024} by exhibiting similar bias reductions. However, the improved estimates for individual parameters are no longer attributable to solely the variable \ce{H2O} profile or the inclusion of clouds. 
    
    We observe two additional important differences between the \vwcf{} and \vwcl{} retrieval results and the previous studies. First, we do not retrieve accurate estimates for \Ps{} and \Ts{} for most views independent of the \R{} and \SN{} of the spectrum. The retrieved \ce{H2O} profiles indicate that the \ce{H2O} abundance in the lower atmosphere is increased by $\geq0.5$~dex compared to the \cwcf{} model. {These elevated \ce{H2O} levels lead to an optically thick lower atmosphere due to the strong MIR continuum opacity of \ce{H2O}. As a result, the high-pressure layers of the model atmosphere do not contribute to the exoplanet's MIR spectrum and therefore only lower limits for \Ps{} and \Ts{} are attainable. The pressure \Popaque{} of the lowest atmosphere layer that still contributes to the modelled MIR spectrum roughly corresponds to the ground truth \Ps{} value. Thus, the model atmosphere becomes opaque to MIR radiation close to Earth's surface.} This agrees well with measurements of the MIR opacity of Earth's atmosphere \citep[see, e.g.,][]{Harries2008}. This finding can be generalized to exoplanets, where clouds and the MIR opacity sources set a fundamental limit on how deep an atmosphere can be probed.

    Second, at the minimal \R{} and \SN{} requirements from \citet{LIFE_III} (\Rv{50}, \SNv{10}), Earth-like \ce{CH4} mass fractions of $10^{-6}$ are not reliably detected in the \vwcf{} and \vwcl{} retrievals. Instead, most of these retrieval runs yield upper limits of $\lgrt{\ce{CH4}}\leq5.0\pm0.5$~dex for the \ce{CH4} abundance. This occurs since Earth's main MIR \ce{CH4} signature at $\sim\mic{8}$ overlaps with a strong \ce{H2O} band (\mic{5-8}). Thus, changes to the \ce{H2O} abundance model directly impact the \ce{CH4} retrieval. Importantly, the main driver for the initial \R{} and \SN{} requirements was the capability to detect \ce{CH4} in an Earth-like atmosphere. Our results indicate that a bias-free constraint on the \ce{CH4} abundance (mean of \ce{CH4} posterior within $0.5$~dex of 1~bar ground truth; uncertainty $\leq\pm1.0$~dex) at \Rv{50} is only possible for $\SN{}\geq20$. For the \Rv{100} scenarios, such \ce{CH4} constraints remain possible for $\SNv{10}$.
    
    Last, while the simulation-based \life{} studies analyzed the MIR emission between \mic{4-18.5}, the empirical Earth spectra studied here and in \citet{Mettler2024} cover a smaller range (\mic{3.75-4.6}, \mic{6.2-15.4}). Yet, the constraints we retrieved for the \cwcf{} model are comparable to the previous studies. This suggests that measurements of the MIR flux above \mic{15.4} and between \mic{4.6-6.2} are not essential to accurately characterize an exo-Earth with \life{}. However, additional studies are required to investigate whether this finding generalizes to arbitrary HZ terrestrial exoplanets.

\subsection{Limitations and Future Work}\label{SubSec:limitations_future_work}

    The present study demonstrates, that adding simple physical constraints to the retrieval forward model can help significantly reduce biases in the retrieved posteriors for an Earth-like spectrum. Yet, there are limitations inherent to our work, the effect of which must be investigated in future studies.

    First, our model for the atmospheric \ce{H2O} profile assumed that condensation occurs if the relative humidity reaches 100\%. Thus, physical tropospheric states, such as the \ce{H2O} supersaturation of air \citep[e.g.,][]{Spichtinger2003,Genthon2017}, are not possible. Yet, since we are only sensitive to order of magnitude changes in the \ce{H2O} abundance for the \R{} and \SN{} scenarios considered, we do not expect such factors to significantly affect our findings. Further, we assumed that only condensation affects the \ce{H2O} profile. However, in the stratosphere and above, the \ce{H2O} abundance is strongly affected by additional processes such as the photochemical oxidation of \ce{CH4} \citep[e.g.,][]{Jones1986,Frank2018}. Yet, since the bulk of Earth's MIR emission originates from the troposphere, we argue that the stratospheric \ce{H2O} profile only affects our characterization negligibly.

    {Second, we assume constant abundance profiles for all trace-gases except for \ce{H2O}. On Earth, especially the \ce{O3} but also the \ce{CH4} abundances exhibit significant altitude dependencies (see Figure~\ref{fig:disk_integrated_input}). These variations are predominantly evoked by photochemical processes in the stratosphere \citep{Chapman1932,Jones1986}. However, in contrast to \ce{H2O}, a simple physical constraint for the \ce{O3} and \ce{CH4} profiles does not exist \citep[e.g.,][]{KozakisO3}. As a result, models for these profiles would be relatively unconstrained. In addition, the MIR spectral feature of \ce{O3} at \mic{9.7} originates from the high pressure layers in Earth's atmosphere ($p>0.3$~bar). Consequentially, we are not sensitive to the large \ce{O3} variations in the upper atmosphere at the considered \R{} and \SN{}. Also for \ce{CH4}, the main contribution to the MIR feature at \mic{7.7} originates from the high pressure layers ($p>10^{-1}$~bar). The vertical variations in the ground truths in these layers ($\pm0.1$~dex) are significantly smaller than the uncertainties on the \ce{CH4} posteriors ($\geq\pm0.5$~dex). Given the limited spectral information available, the impact of the \ce{O3} and \ce{CH4} variations on our results is likely negligible.}

    {Third, instead of modelling the wavelength-dependent MIR opacity of \ce{H2O} clouds, we assume gray clouds. Further, we assume the clouds to form in a single atmosphere layer. In contrast, for Earth, the position of the cloud-top is variable \citep[$\Pcloud{}$ ranges from $\sim0.1-0.9$~bar; e.g.,][]{Kokhanovsky2011}. However, the MIR optical properties of clouds do not vary strongly with wavelength or pressure \citep{Petty2006}. Furthermore, findings presented in \citet{LIFE_IX} suggest that physical cloud properties (such as cloud composition and particle size) cannot be constrained for the \life{} \R{} and \SN{} scenarios considered here. Thus, given the limited information content of the spectra studied, the assumption of gray clouds at a fixed height is justifiable.}

    Fourth, we parameterized the atmospheric \pt{} profile using a fourth order polynomial. This model has the benefit of being highly flexible. Since it does not impose any physical constraints, it can fit any \pt{} structure. However, limiting the atmospheric \pt{} structure to physically viable states by using a learning-based parametrization \citep[see, e.g.][]{TimmyPT2023} could help improve our characterization. This is especially promising for the \vwcf{} and \vwcl{} models, since they link the \ce{H2O} abundance profile and the clouds to the atmospheric \pt{} structure. However, for terrestrial atmospheres, the reliability of learning-based \pt{} models is currently limited by the availability of sufficient training data.

    Fifth, we focus on Earth's MIR emission spectrum, which provides an excellent benchmark \citep[e.g.,][]{Robinson_2023}. Our finding that neglecting vertical abundance variations and clouds affects the retrieved parameter estimates can be generalized to terrestrial exoplanets. Yet, rocky exoplanets will exhibit atmospheres different from Earth. Thus, detailed studies similar to what \citet{Rowland2023} did for cloudy L-dwarfs are required for terrestrial exoplanets to search for additional model assumptions that could affect the interpretation of MIR thermal emission spectra.

    Last, the comparison of different retrieval approaches has revealed that the obtained parameter posteriors can depend on framework particularities such as the radiative transfer implementation, the parameter sampling algorithm, or the used molecular line opacities \citep[e.g.,][]{Line2013,BarstowRetComp,Alei2022}. Community efforts (e.g., the CUISINES Working Group\footnote{\url{https://nexss.info/cuisines/}}) that aim to benchmark, compare, and validate different retrieval frameworks on real and simulated spectral data are indispensable to ensure the correct characterization of HZ terrestrial exoplanet atmospheres in the future.
\section{Summary and Conclusions}\label{sec:conclusions}

    In this study, we investigated how common atmospheric retrieval assumptions can bias the characterization of terrestrial HZ exoplanets. Specifically, we tested how model assumptions for the vertical \ce{H2O} abundance profile and the atmospheric clouds impact our characterization of Earth as an exoplanet by running retrievals with different forward models. The baseline model (\cwcf{}) assumed a cloud-free atmosphere with constant abundance profiles for all trace-gases. While still cloud-free, the \vwcf{} model estimated the \ce{H2O} abundance profile by accounting for \ce{H2O} condensation. Last, the \vwcl{} model allowed for both \ce{H2O} condensation and a partially cloudy atmosphere. We first validated these three models in test retrievals on a simulated 1D low-noise Earth spectrum (\Rv{100}, \SNv{20}). Thereafter, we ran retrievals on \life{} mock observations (\Rv{50,100}; \SNv{10,20}) of empirical disk-integrated MIR thermal emission spectra of Earth representative of different views and seasons. The performance of the three models was benchmarked against {ground truth averages} from remote sensing data.

    {Independent of the assumed forward model, our retrieval results highlight the unique strength of considering an exoplanet's MIR thermal emission. Earth-like \ce{H2O}, \ce{CO2}, and \ce{O3} concentrations are easily constrained to within $\pm1.0$~dex, and \ce{CH4} is reliably detected at \Rv{100}. Further, we significantly constrain the atmospheric \pt{} structure ($\pm0.5$~dex for pressures, $\pm20$~K for temperatures) and the planet radius ($\pm0.1\,R_\oplus$). Yet, our results also demonstrate that simplifying assumptions made by retrieval forward models can bias the characterization. For the \cwcf{} model, pressures are underestimated by $\sim0.5$~dex, and the abundances of \ce{CO2}, \ce{O3}, and \ce{CH4} overestimated by $\geq0.6$~dex. Also, the obtained \Rpl{} estimates are consistently $\geq0.05~R_\oplus$ too low. This translates to biased equilibrium temperature (\Teq{}) and Bond albedo (\Ab{}) estimates (\Teq{} $\geq5$~K too high; \Ab{} $\geq0.05$ too low). With the \vwcf{} forward model, these offsets can already be reduced significantly for most parameters. The \vwcl{} model yields bias-free estimates for the shape of the \pt{} structure, \Rpl{}, the trace-gas abundances, \Teq{}, and \Ab{}. For the surface pressure \Ps{} and temperature \Ts{}, the \vwcf{} and \vwcl{} retrievals yield lower limits. For both models, the high-pressure atmosphere layers are opaque to MIR radiation due to their high \ce{H2O} abundance. Consequentially, these layers do not contribute to the MIR thermal emission spectrum and thus \Ps{} and \Ts{} are not constrainable. These results demonstrate that physically motivated forward models can facilitate the accurate interpretation of spectra.}

    In addition, our work has important implications for the \life{} technical requirements. \citet{LIFE_III} derived preliminary minimal wavelength coverage, \R{}, and \SN{} requirements by running a grid of retrievals on simulated Earth spectra assuming constant abundance profiles and a cloud-free atmosphere. The authors derived lower requirement limits of \mic{4-18.5}, \Rv{50}, and \SNv{10} by requiring a detection of the potential biosignature \ce{CH4}. Yet, our results on empirical Earth spectra suggest, that a bias-free \ce{CH4} detection is not possible for Earth with the current baseline \R{} and \SN{} requirements. Instead, we find that at \SNv{10} a minimal \R{} of 100 is required to ensure a confident and bias-free \ce{CH4} detection. {With a proper treatment of the \ce{H2O} profile and clouds, an \Rv{100} MIR \life{} observation would allow us to correctly identify an exo-Earth as habitable and quantify the abundance of the biosignature gas \ce{CH4}.} Further, our findings suggest that flux measurements above \mic{15.4} and between \mic{4.6-6.2} are not essential for an accurate characterization of an exo-Earth with \life{}. However, this must be confirmed for non Earth-like HZ terrestrial exoplanets.

    First detailed spectral measurements with the JWST have set us off on a journey to characterize and understand the conditions on terrestrial exoplanets. These efforts will intensify with future generations of optimized space missions like \life{} capable of acquiring detailed spectral measurements of nearby temperate terrestrial exoplanets. Inverse modeling approaches such as atmospheric retrievals will be used to analyze these data. Yet, as we demonstrate, simplifying assumptions in retrievals can lead to a biased interpretation. Studies that extend beyond the Earth scenario are essential to build a profound understanding of how different model assumptions may impact the interpretation of observed exoplanet spectra. 
    These efforts are indispensable steps forward in our pursuit of understanding the habitability of distant worlds and are crucial to identifying signs of life outside the Solar System.

\begin{acknowledgments}
This work has been carried out within the framework of the National Centre of Competence in Research PlanetS supported by the Swiss National Science Foundation under grant 51NF40\textunderscore205606. S.P.Q. acknowledges the financial support of the SNSF. B.S.K. acknowledges the support of an ETH Zurich Doc.Mobility Fellowship. E.A.'s research was supported by an appointment to the NASA Postdoctoral Program at the NASA Goddard Space Research, administered by Oak Ridge Associated Universities under contract with NASA.
\end{acknowledgments}

\bibliography{Lit.bib}
\bibliographystyle{aasjournal}

\appendix
\restartappendixnumbering

\section{Supplementary Data from the Retrievals}\label{App:additional_results}

\begin{figure*}
   \centering
    \includegraphics[width=\textwidth]{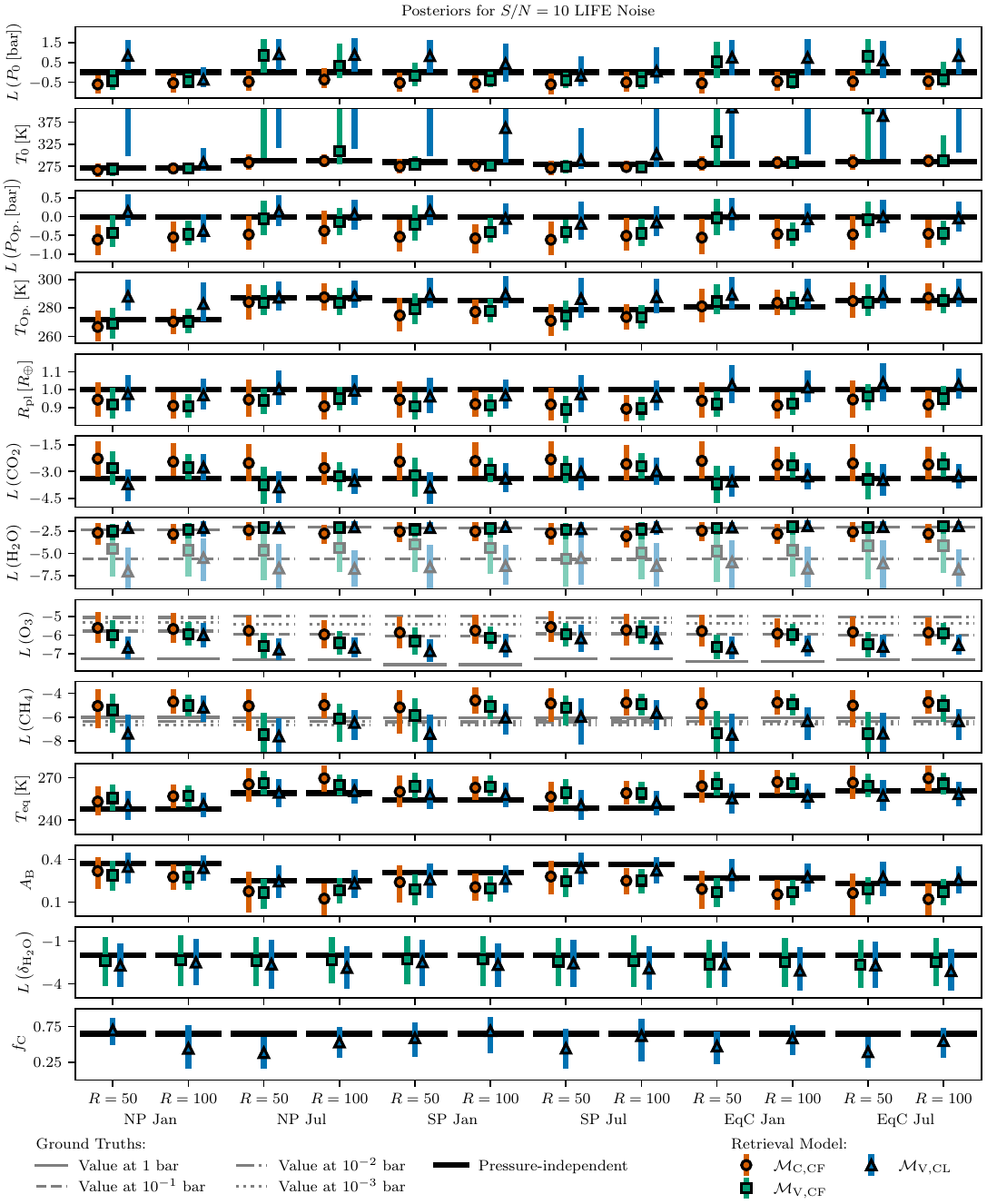}
    \caption{Posteriors from retrievals on the six disk-integrated Earth spectra for the \SNv{10} \lifesim{}-noise scenarios. Here, $L(\cdot)$ stands for $\lgrt{\cdot}$. Thick black lines indicate pressure-independent ground truths. Thin gray lines show the ground truth abundances at different atmospheric pressures (solid: 1~bar; dashed: $10^{-1}$~bar; dashed-dotted: $10^{-2}$~bar; dotted: $10^{-3}$~bar). The markers represent the retrieval models: orange circles -- \cwcf{}, green squares -- \vwcf{}, and blue triangles -- \vwcl{}. The colored lines indicate the $16\%-84\%$ posterior percentiles. For the \vwcf{} and \vwcl{} models, which assume variable \ce{H2O} profiles, we provide both the retrieved \ce{H2O} abundance at the pressure \Popaque{} where the atmosphere becomes optically thick (dark marker color) and at 1~dex below \Popaque{} (light marker color).}
         \label{fig:views_posteriors_SN10}
    \end{figure*}

\begin{figure*}
   \centering
    \includegraphics[width=\textwidth]{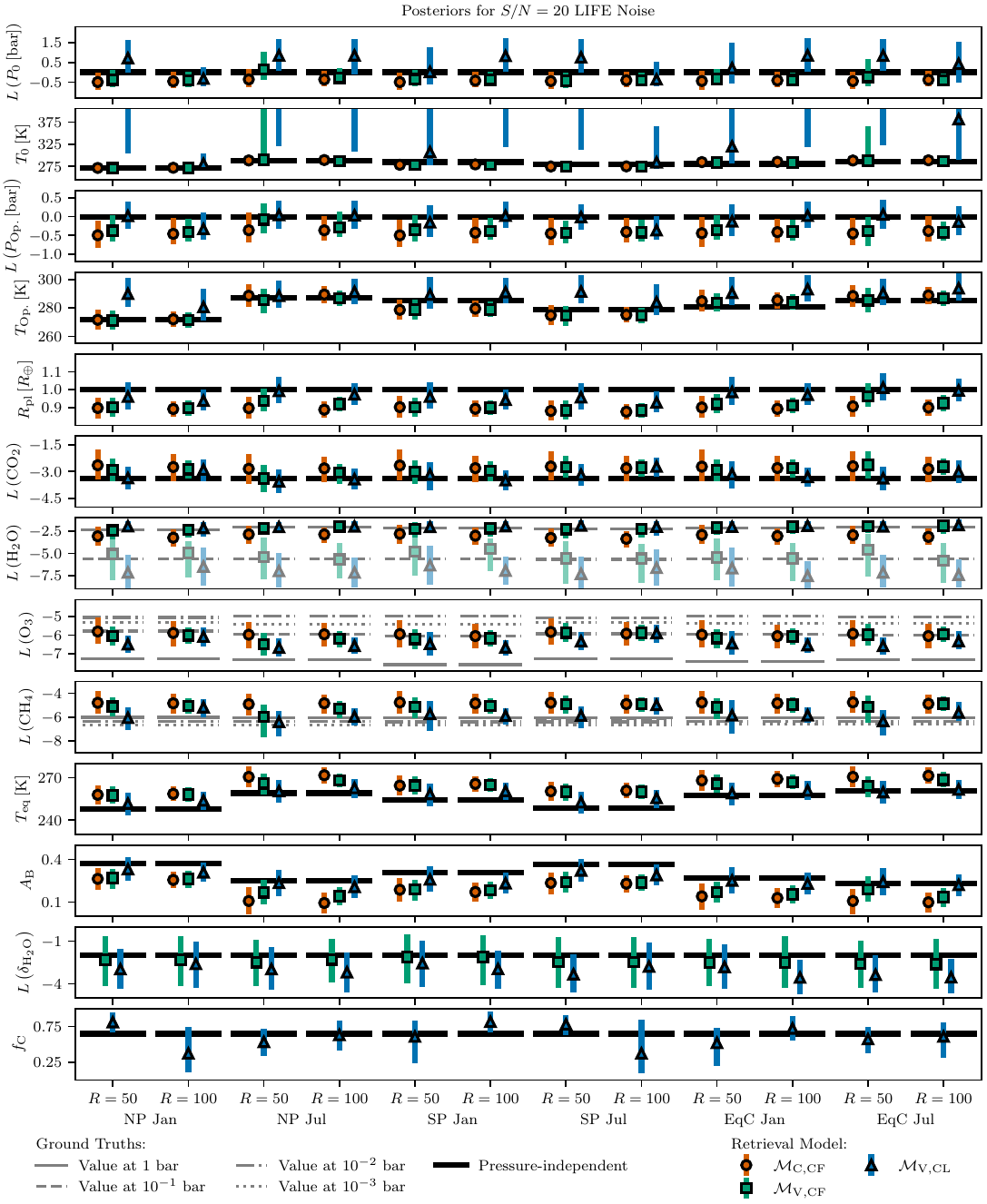}
    \caption{Posteriors from retrievals on the six disk-integrated Earth spectra for the \SNv{20} \lifesim{}-noise scenarios. Here, $L(\cdot)$ stands for $\lgrt{\cdot}$. Thick black lines indicate pressure-independent ground truths. Thin gray lines show the ground truth abundances at different atmospheric pressures (solid: 1~bar; dashed: $10^{-1}$~bar; dashed-dotted: $10^{-2}$~bar; dotted: $10^{-3}$~bar). The markers represent the retrieval models: orange circles -- \cwcf{}, green squares -- \vwcf{}, and blue triangles -- \vwcl{}. The colored lines indicate the $16\%-84\%$ posterior percentiles. For the \vwcf{} and \vwcl{} models, which assume variable \ce{H2O} profiles, we provide both the retrieved \ce{H2O} abundance at the pressure \Popaque{} where the atmosphere becomes optically thick (dark marker color) and at 1~dex below \Popaque{} (light marker color).}
         \label{fig:views_posteriors_SN20}
    \end{figure*}

In Table~\ref{Table:Evidences_and_K} we list the log-evidences $\ln\left(\mathcal{Z}\right)$ for all retrievals and the log-Bayes' factors $\lgrt{K}$ used for model comparison (see Table~\ref{Table:Jeffrey}). {In Figures~\ref{fig:views_posteriors_SN10} and \ref{fig:views_posteriors_SN20}, we present the retrieved posteriors for all considered disk-integrated Earth spectra (Figure~\ref{fig:views_posteriors_SN10}: \SNv{10}; Figure~\ref{fig:views_posteriors_SN20}: \SNv{20}).} In Tables~\ref{tab:Ret_NP} to \ref{tab:Ret_EqC}, we list all the numeric values corresponding to the posteriors in Figures~\ref{fig:views_posteriors_SN10} and \ref{fig:views_posteriors_SN20}:
\begin{itemize}
    \item Table \ref{tab:Ret_NP} $-$ NP viewing angle posteriors,
    \item Table \ref{tab:Ret_SP} $-$ SP viewing angle posteriors,
    \item Table \ref{tab:Ret_EqC} $-$ EqC viewing angle posteriors.
\end{itemize}

\begin{deluxetable}{ccccccccccc}[htb!]
\tablecaption{Evidences $\mathcal{Z}$ for all retrieval runs and the resulting Bayes' factors $K$ for model comparison.}
\label{Table:Evidences_and_K}      
\tablehead{
 & & & &\multicolumn{3}{c}{Evidence $\ln\left(\mathcal{Z}\right)$} &&\multicolumn{3}{c}{Bayes' Factor $\lgrt{K}$}\\\cline{5-7}\cline{9-11}
\colhead{Spectrum} &\colhead{\R{}} &\colhead{\SN{}} & &\cwcf{} &\vwcf{} &\vwcl{} &  &\cwcf{} vs. \vwcf{} &\cwcf{} vs. \vwcl{} &\vwcf{} vs. \vwcl{}
}
\startdata 
Simulated &100 &20 & &$-50.5^{\pm0.1}$ &$-41.9^{\pm0.1}$ &$-40.5^{\pm0.2}$ & &$-3.7^{\pm0.1}$ &$-4.3^{\pm0.1}$ &$-0.6^{\pm0.2}$\\\hline
\multirow{4}{*}{NP Jan} &\multirow{2}{*}{50} &10 & &$-25.5^{\pm0.2}$ &$-24.8^{\pm0.2}$ &$-26.1^{\pm0.2}$ & &$-0.3^{\pm0.2}$ &$0.3^{\pm0.2}$ &$0.3^{\pm0.2}$\\
 & &20 & &$-32.4^{\pm0.2}$ &$-31.9^{\pm0.2}$ &$-32.6^{\pm0.2}$ & &$-0.2^{\pm0.2}$ &$0.1^{\pm0.2}$ &$0.3^{\pm0.2}$\\
 &\multirow{2}{*}{100} &10 & &$-29.4^{\pm0.2}$ &$-29.0^{\pm0.2}$ &$-29.5^{\pm0.2}$ & &$-0.2^{\pm0.2}$ &$0.0^{\pm0.2}$ &$0.2^{\pm0.2}$\\
 & &20 & &$-36.7^{\pm0.2}$ &$-35.7^{\pm0.2}$ &$-35.8^{\pm0.2}$ & &$-0.4^{\pm0.2}$ &$-0.5^{\pm0.2}$ &$0.1^{\pm0.2}$\\\hline
\multirow{4}{*}{NP Jul} &\multirow{2}{*}{50} &10 & &$-25.3^{\pm0.2}$ &$-24.4^{\pm0.2}$ &$-25.5^{\pm0.2}$ & &$-0.2^{\pm0.2}$ &$-0.1^{\pm0.2}$ &$0.2^{\pm0.2}$\\
 & &20 & &$-32.7^{\pm0.2}$ &$-32.8^{\pm0.2}$ &$-32.3^{\pm0.2}$ & &$0.0^{\pm0.2}$ &$-0.2^{\pm0.2}$ &$-0.2^{\pm0.2}$\\
 &\multirow{2}{*}{100} &10 & &$-29.9^{\pm0.2}$ &$-30.7^{\pm0.2}$ &$-29.3^{\pm0.2}$ & &$0.3^{\pm0.2}$ &$-0.1^{\pm0.2}$ &$-0.3^{\pm0.2}$\\
 & &20 & &$-38.4^{\pm0.2}$ &$-38.0^{\pm0.2}$ &$-37.6^{\pm0.2}$ & &$-0.2^{\pm0.2}$ &$-0.4^{\pm0.2}$ &$-0.2^{\pm0.2}$\\\hline
\multirow{4}{*}{SP Jan} &\multirow{2}{*}{50} &10 & &$-25.4^{\pm0.2}$ &$-25.8^{\pm0.2}$ &$-25.6^{\pm0.2}$ & &$0.2^{\pm0.2}$ &$0.1^{\pm0.2}$ &$-0.1^{\pm0.2}$\\
 & &20 & &$-32.2^{\pm0.2}$ &$-31.8^{\pm0.2}$ &$-31.8^{\pm0.2}$ & &$-0.2^{\pm0.2}$ &$-0.2^{\pm0.2}$ &$-0.0^{\pm0.2}$\\
 &\multirow{2}{*}{100} &10 & &$-29.2^{\pm0.2}$ &$-29.1^{\pm0.2}$ &$-29.1^{\pm0.2}$ & &$-0.1^{\pm0.2}$ &$-0.1^{\pm0.2}$ &$0.0^{\pm0.2}$\\
 & &20 & &$-37.1^{\pm0.2}$ &$-36.6^{\pm0.2}$ &$-37.2^{\pm0.2}$ & &$-0.2^{\pm0.2}$ &$0.1^{\pm0.2}$ &$0.2^{\pm0.2}$\\\hline
\multirow{4}{*}{SP Jul} &\multirow{2}{*}{50} &10 & &$-25.9^{\pm0.2}$ &$-25.8^{\pm0.2}$ &$-25.9^{\pm0.2}$ & &$-0.0^{\pm0.2}$ &$0.0^{\pm0.2}$ &$0.0^{\pm0.2}$\\
 & &20 & &$-32.5^{\pm0.2}$ &$-31.8^{\pm0.2}$ &$-33.0^{\pm0.2}$ & &$-0.3^{\pm0.2}$ &$0.2^{\pm0.2}$ &$0.5^{\pm0.2}$\\
 &\multirow{2}{*}{100} &10 & &$-29.6^{\pm0.2}$ &$-29.1^{\pm0.2}$ &$-29.6^{\pm0.2}$ & &$-0.2^{\pm0.2}$ &$0.0^{\pm0.2}$ &$0.2^{\pm0.2}$\\
 & &20 & &$-37.9^{\pm0.2}$ &$-36.8^{\pm0.2}$ &$-37.6^{\pm0.2}$ & &$-0.5^{\pm0.2}$ &$-0.1^{\pm0.2}$ &$0.3^{\pm0.2}$\\\hline
\multirow{4}{*}{EqC Jan} &\multirow{2}{*}{50} &10 & &$-25.4^{\pm0.2}$ &$-25.5^{\pm0.2}$ &$-25.6^{\pm0.2}$ & &$0.0^{\pm0.2}$ &$0.1^{\pm0.2}$ &$-0.0^{\pm0.2}$\\
 & &20 & &$-32.7^{\pm0.2}$ &$-32.8^{\pm0.2}$ &$-32.1^{\pm0.2}$ & &$0.1^{\pm0.2}$ &$-0.3^{\pm0.2}$ &$-0.3^{\pm0.2}$\\
 &\multirow{2}{*}{100} &10 & &$-29.4^{\pm0.2}$ &$-29.4^{\pm0.2}$ &$-29.6^{\pm0.2}$ & &$-0.0^{\pm0.2}$ &$0.1^{\pm0.2}$ &$0.1^{\pm0.2}$\\
 & &20 & &$-38.3^{\pm0.2}$ &$-37.1^{\pm0.2}$ &$-38.0^{\pm0.2}$ & &$-0.5^{\pm0.2}$ &$-0.1^{\pm0.2}$ &$0.4^{\pm0.2}$\\\hline
\multirow{4}{*}{EqC Jul} &\multirow{2}{*}{50} &10 & &$-25.4^{\pm0.2}$ &$-26.2^{\pm0.2}$ &$-25.6^{\pm0.2}$ & &$0.3^{\pm0.2}$ &$0.1^{\pm0.2}$ &$-0.2^{\pm0.2}$\\
 & &20 & &$-33.8^{\pm0.2}$ &$-33.9^{\pm0.2}$ &$-32.6^{\pm0.2}$ & &$0.0^{\pm0.2}$ &$-0.2^{\pm0.2}$ &$-0.3^{\pm0.2}$\\
 &\multirow{2}{*}{100} &10 & &$-29.4^{\pm0.2}$ &$-29.6^{\pm0.2}$ &$-29.9^{\pm0.2}$ & &$0.1^{\pm0.2}$ &$0.2^{\pm0.2}$ &$0.1^{\pm0.2}$\\
 & &20 & &$-38.6^{\pm0.2}$ &$-37.1^{\pm0.2}$ &$-38.4^{\pm0.2}$ & &$-0.6^{\pm0.2}$ &$-0.1^{\pm0.2}$ &$0.6^{\pm0.2}$\\
\enddata
\tablecomments{The $\lgrt{K}$ values were calculated from $\ln\left(\mathcal{Z}\right)$ using Equation~(\ref{eq:BayesFactor}) from Section~\ref{subsec:retrieval_routine}.}
\end{deluxetable}

\movetabledown=68mm
\begin{rotatetable}
\begin{deluxetable}{ccccccccccccccccccccccc}
\tablecaption{Numeric values for the NP view posteriors shown in Figures~\ref{fig:views_posteriors_SN10} and \ref{fig:views_posteriors_SN20}.}
\tablehead{
&&&\multicolumn{4}{c}{Ground Truths} &&\multicolumn{7}{c}{Posteriors for \Rv{50} Spectra} &&\multicolumn{7}{c}{Posteriors for \Rv{100} Spectra} \\ \cline{9-15} \cline{17-23}
&&&\multicolumn{4}{c}{Pressure Levels [bar]} &&\multicolumn{3}{c}{\lifesim{} \SNv{10}}&&\multicolumn{3}{c}{\lifesim{} \SNv{20}} &&\multicolumn{3}{c}{\lifesim{} \SNv{10}}&&\multicolumn{3}{c}{\lifesim{} \SNv{20}} \\\cline{4-7} \cline{9-11} \cline{13-15} \cline{17-19} \cline{21-23}
&\colhead{Parameter} &&\colhead{1}&\colhead{$10^{-1}$}&\colhead{$10^{-2}$}&\colhead{$10^{-3}$} &&\cwcf{}&\vwcf{}&\vwcl{}&&\cwcf{}&\vwcf{}&\vwcl{}&&\cwcf{}&\vwcf{}&\vwcl{}&&\cwcf{}&\vwcf{}&\vwcl{}
}
\startdata
\multirow{13}{*}{\rotatebox[origin=c]{90}{NP Jan View}}
    &$L\left(P_0\left[\mathrm{bar}\right]\right)$ & &\multicolumn{4}{c}{$0.0$} & &$-0.6^{+0.3}_{-0.3}$ &$-0.4^{+0.4}_{-0.3}$ &$0.8^{+0.7}_{-0.6}$ & &$-0.5^{+0.3}_{-0.3}$ &$-0.4^{+0.3}_{-0.2}$ &$0.7^{+0.8}_{-0.6}$ & &$-0.6^{+0.3}_{-0.3}$ &$-0.5^{+0.3}_{-0.2}$ &$-0.4^{+0.5}_{-0.3}$ & &$-0.5^{+0.4}_{-0.2}$ &$-0.4^{+0.3}_{-0.2}$ &$-0.3^{+0.4}_{-0.2}$ \\
    &$L\left(P_\mathrm{Op.}\left[\mathrm{bar}\right]\right)$ & &\multicolumn{4}{c}{$0.0$} & &$-0.6^{+0.3}_{-0.3}$ &$-0.4^{+0.4}_{-0.3}$ &$0.1^{+0.4}_{-0.3}$ & &$-0.5^{+0.3}_{-0.3}$ &$-0.4^{+0.3}_{-0.2}$ &$0.0^{+0.3}_{-0.3}$ & &$-0.6^{+0.3}_{-0.3}$ &$-0.5^{+0.3}_{-0.2}$ &$-0.4^{+0.4}_{-0.3}$ & &$-0.5^{+0.4}_{-0.2}$ &$-0.4^{+0.3}_{-0.2}$ &$-0.3^{+0.4}_{-0.2}$ \\
    &$T_0\left[\mathrm{K}\right]$ & &\multicolumn{4}{c}{$272$} & &$267^{+10}_{-8}$ &$269^{+9}_{-9}$ &$453^{+402}_{-148}$ & &$272^{+5}_{-5}$ &$271^{+5}_{-5}$ &$448^{+465}_{-137}$ & &$270^{+7}_{-7}$ &$270^{+7}_{-6}$ &$283^{+26}_{-11}$ & &$272^{+3}_{-3}$ &$272^{+4}_{-3}$ &$281^{+17}_{-7}$ \\
    &$T_\mathrm{Op.}\left[\mathrm{K}\right]$ & &\multicolumn{4}{c}{$272$} & &$267^{+10}_{-8}$ &$269^{+9}_{-9}$ &$288^{+10}_{-8}$ & &$272^{+5}_{-5}$ &$271^{+5}_{-5}$ &$290^{+10}_{-7}$ & &$270^{+7}_{-7}$ &$270^{+7}_{-6}$ &$283^{+13}_{-10}$ & &$272^{+3}_{-3}$ &$272^{+4}_{-3}$ &$281^{+11}_{-7}$ \\
    &$R_\mathrm{pl}\left[R_\oplus\right]$ & &\multicolumn{4}{c}{$1.0$} & &$0.9^{+0.1}_{-0.1}$ &$0.9^{+0.1}_{-0.1}$ &$1.0^{+0.1}_{-0.1}$ & &$0.9^{+0.0}_{-0.0}$ &$0.9^{+0.0}_{-0.0}$ &$1.0^{+0.1}_{-0.1}$ & &$0.9^{+0.1}_{-0.1}$ &$0.9^{+0.1}_{-0.1}$ &$1.0^{+0.1}_{-0.1}$ & &$0.9^{+0.0}_{-0.0}$ &$0.9^{+0.0}_{-0.0}$ &$0.9^{+0.0}_{-0.0}$ \\
    &$L\left(\mathrm{CO_2}\right)$ & &\multicolumn{4}{c}{$-3.4$} & &$-2.3^{+0.8}_{-0.8}$ &$-2.8^{+0.8}_{-0.8}$ &$-3.7^{+0.7}_{-0.8}$ & &$-2.6^{+0.7}_{-0.6}$ &$-2.9^{+0.5}_{-0.5}$ &$-3.4^{+0.5}_{-0.5}$ & &$-2.5^{+0.9}_{-0.6}$ &$-2.8^{+0.6}_{-0.5}$ &$-2.8^{+0.6}_{-0.6}$ & &$-2.7^{+0.6}_{-0.6}$ &$-2.9^{+0.3}_{-0.4}$ &$-2.9^{+0.4}_{-0.5}$ \\
    &$L\left(\mathrm{O_3}\right)$ & &$-7.3$ &$-5.8$ &$-5.0$ &$-5.3$ & &$-5.6^{+0.7}_{-0.7}$ &$-6.0^{+0.6}_{-0.6}$ &$-6.7^{+0.4}_{-0.5}$ & &$-5.8^{+0.6}_{-0.5}$ &$-6.0^{+0.4}_{-0.4}$ &$-6.5^{+0.3}_{-0.3}$ & &$-5.7^{+0.7}_{-0.6}$ &$-5.9^{+0.5}_{-0.4}$ &$-6.0^{+0.5}_{-0.5}$ & &$-5.9^{+0.5}_{-0.5}$ &$-6.0^{+0.3}_{-0.3}$ &$-6.1^{+0.3}_{-0.4}$ \\
    &$L\left(\mathrm{CH_4}\right)$ & &$-6.0$ &$-6.1$ &$-6.4$ &$-6.7$ & &$-5.1^{+1.2}_{-1.7}$ &$-5.4^{+1.1}_{-1.7}$ &$-7.4^{+1.3}_{-1.5}$ & &$-4.8^{+0.7}_{-0.7}$ &$-5.1^{+0.5}_{-0.6}$ &$-6.1^{+0.6}_{-0.8}$ & &$-4.7^{+0.9}_{-0.8}$ &$-5.0^{+0.7}_{-0.7}$ &$-5.2^{+0.7}_{-1.0}$ & &$-4.8^{+0.6}_{-0.7}$ &$-5.0^{+0.4}_{-0.4}$ &$-5.2^{+0.5}_{-0.6}$ \\
    &$L\left(\mathrm{H_2O}\right)$ & &$-2.3$ &$-5.6$ &... &... & &$-2.7^{+0.8}_{-1.0}$ &$-2.5^{+0.5}_{-0.7}$ &$-2.1^{+0.3}_{-0.4}$ & &$-3.1^{+0.8}_{-0.7}$ &$-2.4^{+0.4}_{-0.7}$ &$-2.0^{+0.3}_{-0.3}$ & &$-2.8^{+0.8}_{-0.8}$ &$-2.4^{+0.4}_{-0.6}$ &$-2.1^{+0.4}_{-0.7}$ & &$-3.2^{+0.6}_{-0.7}$ &$-2.3^{+0.3}_{-0.6}$ &$-2.1^{+0.4}_{-0.7}$ \\
    &$L\left(\delta_\mathrm{H_2O}\right)$ & &\multicolumn{4}{c}{$-2.0$} & &... &$-2.4^{+1.5}_{-1.6}$ &$-2.7^{+1.4}_{-1.3}$ & &... &$-2.3^{+1.5}_{-1.7}$ &$-3.0^{+1.2}_{-1.2}$ & &... &$-2.3^{+1.5}_{-1.6}$ &$-2.5^{+1.4}_{-1.4}$ & &... &$-2.3^{+1.5}_{-1.6}$ &$-2.6^{+1.4}_{-1.5}$ \\
    &$f_\mathrm{C}$ & &\multicolumn{4}{c}{$0.7$} & &... &... &$0.7^{+0.1}_{-0.2}$ & &... &... &$0.8^{+0.1}_{-0.1}$ & &... &... &$0.4^{+0.3}_{-0.2}$ & &... &... &$0.4^{+0.3}_{-0.2}$ \\
    &$T_\mathrm{eq}\left[\mathrm{K}\right]$ & &\multicolumn{4}{c}{$248$} & &$253^{+9}_{-8}$ &$256^{+8}_{-7}$ &$250^{+9}_{-8}$ & &$258^{+5}_{-5}$ &$257^{+4}_{-4}$ &$252^{+6}_{-6}$ & &$257^{+6}_{-6}$ &$257^{+6}_{-6}$ &$251^{+6}_{-7}$ & &$259^{+3}_{-3}$ &$258^{+3}_{-3}$ &$254^{+4}_{-4}$ \\
    &$A_\mathrm{B}$ & &\multicolumn{4}{c}{$0.37$} & &$0.32^{+0.08}_{-0.10}$ &$0.29^{+0.08}_{-0.09}$ &$0.35^{+0.08}_{-0.10}$ & &$0.26^{+0.05}_{-0.06}$ &$0.27^{+0.05}_{-0.05}$ &$0.33^{+0.06}_{-0.06}$ & &$0.28^{+0.07}_{-0.07}$ &$0.27^{+0.06}_{-0.07}$ &$0.34^{+0.07}_{-0.07}$ & &$0.26^{+0.04}_{-0.04}$ &$0.26^{+0.04}_{-0.04}$ &$0.31^{+0.05}_{-0.05}$ \\\hline
\multirow{13}{*}{\rotatebox[origin=c]{90}{NP Jul View}}
    &$L\left(P_0\left[\mathrm{bar}\right]\right)$ & &\multicolumn{4}{c}{$0.0$} & &$-0.5^{+0.4}_{-0.3}$ &$0.8^{+0.7}_{-0.7}$ &$0.9^{+0.6}_{-0.6}$ & &$-0.4^{+0.4}_{-0.2}$ &$0.1^{+0.8}_{-0.4}$ &$0.8^{+0.7}_{-0.6}$ & &$-0.4^{+0.5}_{-0.3}$ &$0.3^{+1.0}_{-0.5}$ &$0.9^{+0.7}_{-0.7}$ & &$-0.4^{+0.3}_{-0.2}$ &$-0.3^{+0.4}_{-0.2}$ &$0.8^{+0.7}_{-0.8}$ \\
    &$L\left(P_\mathrm{Op.}\left[\mathrm{bar}\right]\right)$ & &\multicolumn{4}{c}{$0.0$} & &$-0.5^{+0.4}_{-0.3}$ &$-0.1^{+0.4}_{-0.4}$ &$0.1^{+0.3}_{-0.3}$ & &$-0.4^{+0.4}_{-0.2}$ &$-0.1^{+0.3}_{-0.3}$ &$0.0^{+0.3}_{-0.3}$ & &$-0.4^{+0.5}_{-0.3}$ &$-0.1^{+0.3}_{-0.3}$ &$0.1^{+0.3}_{-0.3}$ & &$-0.4^{+0.3}_{-0.2}$ &$-0.3^{+0.4}_{-0.2}$ &$0.0^{+0.3}_{-0.3}$ \\
    &$T_0\left[\mathrm{K}\right]$ & &\multicolumn{4}{c}{$287$} & &$284^{+11}_{-10}$ &$440^{+406}_{-140}$ &$477^{+403}_{-154}$ & &$289^{+6}_{-6}$ &$290^{+154}_{-6}$ &$484^{+405}_{-156}$ & &$287^{+8}_{-7}$ &$310^{+281}_{-25}$ &$482^{+433}_{-161}$ & &$289^{+4}_{-4}$ &$287^{+3}_{-3}$ &$487^{+456}_{-174}$ \\
    &$T_\mathrm{Op.}\left[\mathrm{K}\right]$ & &\multicolumn{4}{c}{$287$} & &$284^{+11}_{-10}$ &$284^{+10}_{-6}$ &$287^{+9}_{-7}$ & &$289^{+6}_{-6}$ &$286^{+6}_{-7}$ &$288^{+8}_{-6}$ & &$287^{+8}_{-7}$ &$284^{+9}_{-7}$ &$289^{+8}_{-7}$ & &$289^{+4}_{-4}$ &$287^{+3}_{-3}$ &$291^{+8}_{-6}$ \\
    &$R_\mathrm{pl}\left[R_\oplus\right]$ & &\multicolumn{4}{c}{$1.0$} & &$0.9^{+0.1}_{-0.1}$ &$0.9^{+0.1}_{-0.1}$ &$1.0^{+0.1}_{-0.1}$ & &$0.9^{+0.0}_{-0.0}$ &$0.9^{+0.1}_{-0.0}$ &$1.0^{+0.1}_{-0.1}$ & &$0.9^{+0.1}_{-0.1}$ &$0.9^{+0.0}_{-0.0}$ &$1.0^{+0.1}_{-0.1}$ & &$0.9^{+0.0}_{-0.0}$ &$0.9^{+0.0}_{-0.0}$ &$1.0^{+0.0}_{-0.0}$ \\
    &$L\left(\mathrm{CO_2}\right)$ & &\multicolumn{4}{c}{$-3.4$} & &$-2.5^{+0.9}_{-0.9}$ &$-3.7^{+0.9}_{-0.9}$ &$-3.9^{+0.7}_{-0.7}$ & &$-2.8^{+0.6}_{-0.7}$ &$-3.4^{+0.6}_{-0.6}$ &$-3.6^{+0.5}_{-0.5}$ & &$-2.8^{+0.8}_{-0.8}$ &$-3.3^{+0.6}_{-0.7}$ &$-3.5^{+0.6}_{-0.6}$ & &$-2.8^{+0.5}_{-0.6}$ &$-3.0^{+0.3}_{-0.5}$ &$-3.4^{+0.4}_{-0.4}$ \\
    &$L\left(\mathrm{O_3}\right)$ & &$-7.3$ &$-6.0$ &$-5.0$ &$-5.4$ & &$-5.8^{+0.7}_{-0.6}$ &$-6.6^{+0.5}_{-0.5}$ &$-6.8^{+0.4}_{-0.4}$ & &$-6.0^{+0.5}_{-0.6}$ &$-6.5^{+0.4}_{-0.5}$ &$-6.7^{+0.3}_{-0.3}$ & &$-6.0^{+0.6}_{-0.6}$ &$-6.4^{+0.5}_{-0.5}$ &$-6.7^{+0.4}_{-0.4}$ & &$-6.0^{+0.5}_{-0.5}$ &$-6.2^{+0.3}_{-0.3}$ &$-6.6^{+0.3}_{-0.3}$ \\
    &$L\left(\mathrm{CH_4}\right)$ & &$-6.0$ &$-6.1$ &$-6.4$ &$-6.7$ & &$-5.1^{+1.2}_{-1.8}$ &$-7.5^{+1.6}_{-1.5}$ &$-7.6^{+1.3}_{-1.4}$ & &$-4.9^{+0.6}_{-0.8}$ &$-6.0^{+0.8}_{-1.4}$ &$-6.4^{+0.7}_{-1.0}$ & &$-5.0^{+0.8}_{-0.9}$ &$-6.2^{+1.0}_{-1.7}$ &$-6.5^{+0.8}_{-1.2}$ & &$-4.8^{+0.5}_{-0.6}$ &$-5.3^{+0.4}_{-0.6}$ &$-6.0^{+0.5}_{-0.5}$ \\
    &$L\left(\mathrm{H_2O}\right)$ & &$-2.1$ &$-5.6$ &... &... & &$-2.4^{+0.6}_{-0.8}$ &$-2.1^{+0.3}_{-0.4}$ &$-2.2^{+0.3}_{-0.3}$ & &$-2.9^{+0.7}_{-0.7}$ &$-2.2^{+0.3}_{-0.4}$ &$-2.0^{+0.3}_{-0.3}$ & &$-2.8^{+0.7}_{-0.8}$ &$-2.1^{+0.3}_{-0.4}$ &$-2.0^{+0.3}_{-0.3}$ & &$-2.9^{+0.5}_{-0.6}$ &$-2.0^{+0.3}_{-0.4}$ &$-2.0^{+0.3}_{-0.2}$ \\
    &$L\left(\delta_\mathrm{H_2O}\right)$ & &\multicolumn{4}{c}{$-2.0$} & &... &$-2.4^{+1.5}_{-1.6}$ &$-2.6^{+1.5}_{-1.5}$ & &... &$-2.5^{+1.4}_{-1.5}$ &$-3.0^{+1.3}_{-1.2}$ & &... &$-2.3^{+1.4}_{-1.4}$ &$-2.9^{+1.3}_{-1.3}$ & &... &$-2.3^{+1.2}_{-1.4}$ &$-3.2^{+1.2}_{-1.2}$ \\
    &$f_\mathrm{C}$ & &\multicolumn{4}{c}{$0.7$} & &... &... &$0.4^{+0.2}_{-0.2}$ & &... &... &$0.5^{+0.1}_{-0.2}$ & &... &... &$0.5^{+0.2}_{-0.2}$ & &... &... &$0.6^{+0.2}_{-0.2}$ \\
    &$T_\mathrm{eq}\left[\mathrm{K}\right]$ & &\multicolumn{4}{c}{$259$} & &$265^{+10}_{-10}$ &$266^{+7}_{-6}$ &$259^{+8}_{-8}$ & &$270^{+5}_{-5}$ &$266^{+5}_{-6}$ &$260^{+6}_{-6}$ & &$269^{+7}_{-7}$ &$265^{+6}_{-5}$ &$261^{+7}_{-7}$ & &$272^{+4}_{-4}$ &$268^{+3}_{-3}$ &$263^{+4}_{-5}$ \\
    &$A_\mathrm{B}$ & &\multicolumn{4}{c}{$0.25$} & &$0.18^{+0.12}_{-0.13}$ &$0.17^{+0.07}_{-0.09}$ &$0.25^{+0.09}_{-0.09}$ & &$0.11^{+0.08}_{-0.07}$ &$0.17^{+0.07}_{-0.06}$ &$0.23^{+0.07}_{-0.07}$ & &$0.12^{+0.09}_{-0.10}$ &$0.18^{+0.07}_{-0.07}$ &$0.23^{+0.08}_{-0.08}$ & &$0.09^{+0.05}_{-0.05}$ &$0.14^{+0.04}_{-0.05}$ &$0.21^{+0.06}_{-0.06}$
\enddata
\addtolength{\leftskip} {-0cm}
\addtolength{\rightskip}{-7.3cm}
\tablecomments{Here, $L(\cdot)$ stands for $\lgrt{\cdot}$. In columns two to five, we list the ground truth values for the view. If independent of the atmospheric pressure, we provide a single value. Otherwise, we provide ground truth values at $1$~bar, $10^{-1}$~bar, $10^{-2}$~bar, and $10^{-3}$~bar where available. In the last twelve columns, we list the median and the $16\% - 84\%$ range (via $+/-$ indices) of the parameter posteriors for all combinations of \R{} (50, 100), \SN{} (10, 20), and retrieval model (\cwcf{}, \vwcf{}, \vwcl{}). For \ce{H2O}, all listed posterior values correspond to the abundances at the pressure \Popaque{} where the atmosphere becomes optically thick.}
\label{tab:Ret_NP}
\end{deluxetable}
\end{rotatetable}

\movetabledown=68mm
\begin{rotatetable}
\begin{deluxetable}{ccccccccccccccccccccccc}
\tablecaption{Numeric values for the SP view posteriors shown in Figures~\ref{fig:views_posteriors_SN10} and \ref{fig:views_posteriors_SN20}.}
\tablehead{
&&&\multicolumn{4}{c}{Ground Truths} &&\multicolumn{7}{c}{Posteriors for \Rv{50} Spectra} &&\multicolumn{7}{c}{Posteriors for \Rv{100} Spectra} \\ \cline{9-15} \cline{17-23}
&&&\multicolumn{4}{c}{Pressure Levels [bar]} &&\multicolumn{3}{c}{\lifesim{} \SNv{10}}&&\multicolumn{3}{c}{\lifesim{} \SNv{20}} &&\multicolumn{3}{c}{\lifesim{} \SNv{10}}&&\multicolumn{3}{c}{\lifesim{} \SNv{20}} \\\cline{4-7} \cline{9-11} \cline{13-15} \cline{17-19} \cline{21-23}
&\colhead{Parameter} &&\colhead{1}&\colhead{$10^{-1}$}&\colhead{$10^{-2}$}&\colhead{$10^{-3}$} &&\cwcf{}&\vwcf{}&\vwcl{}&&\cwcf{}&\vwcf{}&\vwcl{}&&\cwcf{}&\vwcf{}&\vwcl{}&&\cwcf{}&\vwcf{}&\vwcl{}
}
\startdata
\multirow{13}{*}{\rotatebox[origin=c]{90}{SP Jan View}}
    &$L\left(P_0\left[\mathrm{bar}\right]\right)$ & &\multicolumn{4}{c}{$0.0$} & &$-0.5^{+0.4}_{-0.3}$ &$-0.2^{+0.5}_{-0.4}$ &$0.8^{+0.7}_{-0.7}$ & &$-0.5^{+0.3}_{-0.2}$ &$-0.4^{+0.3}_{-0.2}$ &$0.0^{+1.1}_{-0.5}$ & &$-0.6^{+0.3}_{-0.3}$ &$-0.4^{+0.3}_{-0.2}$ &$0.4^{+0.9}_{-0.8}$ & &$-0.4^{+0.3}_{-0.2}$ &$-0.4^{+0.3}_{-0.2}$ &$0.8^{+0.7}_{-0.7}$ \\
    &$L\left(P_\mathrm{Op.}\left[\mathrm{bar}\right]\right)$ & &\multicolumn{4}{c}{$0.0$} & &$-0.5^{+0.4}_{-0.3}$ &$-0.2^{+0.4}_{-0.4}$ &$0.2^{+0.3}_{-0.3}$ & &$-0.5^{+0.3}_{-0.2}$ &$-0.4^{+0.3}_{-0.2}$ &$-0.2^{+0.4}_{-0.3}$ & &$-0.6^{+0.3}_{-0.3}$ &$-0.4^{+0.3}_{-0.2}$ &$-0.1^{+0.3}_{-0.3}$ & &$-0.4^{+0.3}_{-0.2}$ &$-0.4^{+0.3}_{-0.2}$ &$0.0^{+0.3}_{-0.3}$ \\
    &$T_0\left[\mathrm{K}\right]$ & &\multicolumn{4}{c}{$285$} & &$275^{+10}_{-9}$ &$279^{+13}_{-8}$ &$439^{+369}_{-133}$ & &$279^{+5}_{-5}$ &$279^{+5}_{-5}$ &$307^{+297}_{-23}$ & &$277^{+7}_{-7}$ &$278^{+7}_{-6}$ &$362^{+407}_{-72}$ & &$279^{+4}_{-4}$ &$279^{+3}_{-3}$ &$504^{+513}_{-179}$ \\
    &$T_\mathrm{Op.}\left[\mathrm{K}\right]$ & &\multicolumn{4}{c}{$285$} & &$275^{+11}_{-9}$ &$279^{+9}_{-8}$ &$289^{+10}_{-7}$ & &$279^{+5}_{-5}$ &$279^{+5}_{-5}$ &$289^{+11}_{-8}$ & &$277^{+7}_{-7}$ &$278^{+7}_{-6}$ &$290^{+11}_{-8}$ & &$279^{+4}_{-4}$ &$279^{+3}_{-3}$ &$291^{+8}_{-6}$ \\
    &$R_\mathrm{pl}\left[R_\oplus\right]$ & &\multicolumn{4}{c}{$1.0$} & &$0.9^{+0.1}_{-0.1}$ &$0.9^{+0.1}_{-0.1}$ &$1.0^{+0.1}_{-0.1}$ & &$0.9^{+0.0}_{-0.0}$ &$0.9^{+0.0}_{-0.0}$ &$1.0^{+0.1}_{-0.0}$ & &$0.9^{+0.1}_{-0.1}$ &$0.9^{+0.0}_{-0.0}$ &$1.0^{+0.1}_{-0.1}$ & &$0.9^{+0.0}_{-0.0}$ &$0.9^{+0.0}_{-0.0}$ &$0.9^{+0.0}_{-0.0}$ \\
    &$L\left(\mathrm{CO_2}\right)$ & &\multicolumn{4}{c}{$-3.4$} & &$-2.4^{+0.9}_{-0.9}$ &$-3.2^{+0.8}_{-1.0}$ &$-3.9^{+0.7}_{-0.8}$ & &$-2.7^{+0.7}_{-0.7}$ &$-3.0^{+0.5}_{-0.5}$ &$-3.2^{+0.6}_{-0.7}$ & &$-2.4^{+0.9}_{-0.7}$ &$-2.9^{+0.5}_{-0.6}$ &$-3.4^{+0.7}_{-0.6}$ & &$-2.8^{+0.6}_{-0.6}$ &$-2.9^{+0.4}_{-0.4}$ &$-3.5^{+0.4}_{-0.4}$ \\
    &$L\left(\mathrm{O_3}\right)$ & &$-7.6$ &$-6.0$ &$-5.0$ &$-5.4$ & &$-5.9^{+0.7}_{-0.7}$ &$-6.3^{+0.6}_{-0.6}$ &$-6.8^{+0.5}_{-0.4}$ & &$-5.9^{+0.6}_{-0.6}$ &$-6.2^{+0.4}_{-0.4}$ &$-6.5^{+0.5}_{-0.5}$ & &$-5.8^{+0.7}_{-0.6}$ &$-6.1^{+0.4}_{-0.4}$ &$-6.6^{+0.5}_{-0.4}$ & &$-6.0^{+0.5}_{-0.5}$ &$-6.2^{+0.3}_{-0.3}$ &$-6.7^{+0.3}_{-0.3}$ \\
    &$L\left(\mathrm{CH_4}\right)$ & &$-6.1$ &$-6.1$ &$-6.4$ &$-6.7$ & &$-5.2^{+1.2}_{-2.0}$ &$-5.9^{+1.3}_{-2.0}$ &$-7.4^{+1.4}_{-1.6}$ & &$-4.8^{+0.7}_{-0.7}$ &$-5.1^{+0.6}_{-0.7}$ &$-5.7^{+0.9}_{-1.2}$ & &$-4.6^{+0.9}_{-0.8}$ &$-5.1^{+0.7}_{-0.8}$ &$-6.0^{+0.9}_{-1.2}$ & &$-4.8^{+0.5}_{-0.6}$ &$-5.1^{+0.4}_{-0.5}$ &$-5.9^{+0.4}_{-0.5}$ \\
    &$L\left(\mathrm{H_2O}\right)$ & &$-2.1$ &$-5.7$ &... &... & &$-2.5^{+0.8}_{-0.9}$ &$-2.3^{+0.5}_{-0.5}$ &$-2.1^{+0.3}_{-0.4}$ & &$-2.8^{+0.7}_{-0.8}$ &$-2.2^{+0.4}_{-0.6}$ &$-2.0^{+0.3}_{-0.4}$ & &$-2.5^{+0.8}_{-0.7}$ &$-2.2^{+0.4}_{-0.7}$ &$-2.0^{+0.3}_{-0.4}$ & &$-3.0^{+0.6}_{-0.7}$ &$-2.2^{+0.4}_{-0.6}$ &$-1.9^{+0.2}_{-0.3}$ \\
    &$L\left(\delta_\mathrm{H_2O}\right)$ & &\multicolumn{4}{c}{$-2.0$} & &... &$-2.3^{+1.4}_{-1.6}$ &$-2.5^{+1.4}_{-1.5}$ & &... &$-2.1^{+1.4}_{-1.6}$ &$-2.6^{+1.4}_{-1.4}$ & &... &$-2.3^{+1.4}_{-1.7}$ &$-2.7^{+1.3}_{-1.4}$ & &... &$-2.1^{+1.4}_{-1.8}$ &$-3.0^{+1.1}_{-1.2}$ \\
    &$f_\mathrm{C}$ & &\multicolumn{4}{c}{$0.7$} & &... &... &$0.6^{+0.2}_{-0.2}$ & &... &... &$0.6^{+0.2}_{-0.3}$ & &... &... &$0.7^{+0.2}_{-0.3}$ & &... &... &$0.8^{+0.1}_{-0.1}$ \\
    &$T_\mathrm{eq}\left[\mathrm{K}\right]$ & &\multicolumn{4}{c}{$254$} & &$260^{+10}_{-9}$ &$264^{+7}_{-7}$ &$258^{+9}_{-8}$ & &$264^{+5}_{-5}$ &$264^{+4}_{-4}$ &$258^{+5}_{-6}$ & &$263^{+6}_{-7}$ &$264^{+6}_{-5}$ &$258^{+6}_{-7}$ & &$266^{+3}_{-3}$ &$265^{+3}_{-3}$ &$260^{+4}_{-4}$ \\
    &$A_\mathrm{B}$ & &\multicolumn{4}{c}{$0.31$} & &$0.24^{+0.10}_{-0.12}$ &$0.19^{+0.08}_{-0.09}$ &$0.26^{+0.09}_{-0.11}$ & &$0.19^{+0.06}_{-0.07}$ &$0.19^{+0.05}_{-0.06}$ &$0.26^{+0.07}_{-0.07}$ & &$0.20^{+0.08}_{-0.08}$ &$0.20^{+0.06}_{-0.07}$ &$0.26^{+0.08}_{-0.08}$ & &$0.17^{+0.04}_{-0.05}$ &$0.18^{+0.04}_{-0.04}$ &$0.23^{+0.05}_{-0.05}$ \\\hline
\multirow{13}{*}{\rotatebox[origin=c]{90}{SP Jul View}}
    &$L\left(P_0\left[\mathrm{bar}\right]\right)$ & &\multicolumn{4}{c}{$0.0$} & &$-0.6^{+0.4}_{-0.3}$ &$-0.4^{+0.3}_{-0.2}$ &$-0.2^{+0.8}_{-0.4}$ & &$-0.4^{+0.3}_{-0.2}$ &$-0.4^{+0.3}_{-0.2}$ &$0.8^{+0.8}_{-0.6}$ & &$-0.5^{+0.4}_{-0.3}$ &$-0.4^{+0.3}_{-0.2}$ &$0.0^{+1.1}_{-0.5}$ & &$-0.4^{+0.3}_{-0.2}$ &$-0.4^{+0.3}_{-0.2}$ &$-0.4^{+0.8}_{-0.2}$ \\
    &$L\left(P_\mathrm{Op.}\left[\mathrm{bar}\right]\right)$ & &\multicolumn{4}{c}{$0.0$} & &$-0.6^{+0.4}_{-0.3}$ &$-0.4^{+0.3}_{-0.2}$ &$-0.2^{+0.5}_{-0.4}$ & &$-0.4^{+0.3}_{-0.2}$ &$-0.4^{+0.3}_{-0.2}$ &$-0.0^{+0.3}_{-0.2}$ & &$-0.5^{+0.4}_{-0.3}$ &$-0.4^{+0.3}_{-0.2}$ &$-0.2^{+0.4}_{-0.3}$ & &$-0.4^{+0.3}_{-0.2}$ &$-0.4^{+0.3}_{-0.2}$ &$-0.4^{+0.3}_{-0.2}$ \\
    &$T_0\left[\mathrm{K}\right]$ & &\multicolumn{4}{c}{$279$} & &$271^{+10}_{-9}$ &$274^{+9}_{-8}$ &$289^{+65}_{-14}$ & &$275^{+5}_{-5}$ &$275^{+5}_{-5}$ &$480^{+477}_{-161}$ & &$274^{+7}_{-7}$ &$273^{+7}_{-6}$ &$303^{+299}_{-23}$ & &$275^{+4}_{-4}$ &$275^{+3}_{-3}$ &$284^{+75}_{-8}$ \\
    &$T_\mathrm{Op.}\left[\mathrm{K}\right]$ & &\multicolumn{4}{c}{$279$} & &$271^{+10}_{-9}$ &$274^{+9}_{-8}$ &$286^{+13}_{-11}$ & &$275^{+5}_{-5}$ &$275^{+5}_{-5}$ &$291^{+10}_{-6}$ & &$274^{+7}_{-7}$ &$273^{+7}_{-6}$ &$288^{+11}_{-9}$ & &$275^{+4}_{-4}$ &$275^{+3}_{-3}$ &$284^{+11}_{-7}$ \\
    &$R_\mathrm{pl}\left[R_\oplus\right]$ & &\multicolumn{4}{c}{$1.0$} & &$0.9^{+0.1}_{-0.1}$ &$0.9^{+0.1}_{-0.1}$ &$1.0^{+0.1}_{-0.1}$ & &$0.9^{+0.0}_{-0.0}$ &$0.9^{+0.0}_{-0.0}$ &$1.0^{+0.1}_{-0.1}$ & &$0.9^{+0.1}_{-0.1}$ &$0.9^{+0.0}_{-0.1}$ &$1.0^{+0.1}_{-0.1}$ & &$0.9^{+0.0}_{-0.0}$ &$0.9^{+0.0}_{-0.0}$ &$0.9^{+0.0}_{-0.0}$ \\
    &$L\left(\mathrm{CO_2}\right)$ & &\multicolumn{4}{c}{$-3.4$} & &$-2.3^{+0.9}_{-0.9}$ &$-2.9^{+0.6}_{-0.6}$ &$-3.0^{+0.7}_{-0.8}$ & &$-2.7^{+0.7}_{-0.6}$ &$-2.7^{+0.5}_{-0.4}$ &$-3.2^{+0.4}_{-0.5}$ & &$-2.6^{+0.9}_{-0.8}$ &$-2.7^{+0.6}_{-0.5}$ &$-3.0^{+0.6}_{-0.6}$ & &$-2.8^{+0.5}_{-0.6}$ &$-2.8^{+0.3}_{-0.4}$ &$-2.7^{+0.4}_{-0.4}$ \\
    &$L\left(\mathrm{O_3}\right)$ & &$-7.3$ &$-5.9$ &$-5.1$ &$-5.3$ & &$-5.6^{+0.7}_{-0.7}$ &$-6.0^{+0.5}_{-0.5}$ &$-6.2^{+0.6}_{-0.6}$ & &$-5.8^{+0.6}_{-0.5}$ &$-5.9^{+0.4}_{-0.4}$ &$-6.3^{+0.3}_{-0.3}$ & &$-5.7^{+0.7}_{-0.7}$ &$-5.8^{+0.5}_{-0.5}$ &$-6.2^{+0.5}_{-0.5}$ & &$-5.9^{+0.5}_{-0.5}$ &$-5.9^{+0.3}_{-0.3}$ &$-5.9^{+0.3}_{-0.4}$ \\
    &$L\left(\mathrm{CH_4}\right)$ & &$-6.1$ &$-6.1$ &$-6.4$ &$-6.7$ & &$-4.9^{+1.1}_{-1.3}$ &$-5.2^{+0.8}_{-1.3}$ &$-6.0^{+1.3}_{-2.1}$ & &$-4.8^{+0.7}_{-0.7}$ &$-4.9^{+0.5}_{-0.6}$ &$-5.9^{+0.6}_{-0.8}$ & &$-4.8^{+0.9}_{-0.8}$ &$-4.9^{+0.7}_{-0.7}$ &$-5.6^{+0.8}_{-1.2}$ & &$-4.9^{+0.5}_{-0.6}$ &$-4.9^{+0.4}_{-0.4}$ &$-5.0^{+0.4}_{-0.6}$ \\
    &$L\left(\mathrm{H_2O}\right)$ & &$-2.2$ &$-5.7$ &... &... & &$-2.7^{+0.8}_{-1.0}$ &$-2.3^{+0.4}_{-0.7}$ &$-2.2^{+0.5}_{-0.7}$ & &$-3.3^{+0.7}_{-0.7}$ &$-2.3^{+0.4}_{-0.7}$ &$-1.9^{+0.2}_{-0.3}$ & &$-3.1^{+0.9}_{-0.9}$ &$-2.3^{+0.5}_{-0.8}$ &$-2.0^{+0.3}_{-0.5}$ & &$-3.4^{+0.6}_{-0.6}$ &$-2.2^{+0.4}_{-0.7}$ &$-2.0^{+0.3}_{-0.7}$ \\
    &$L\left(\delta_\mathrm{H_2O}\right)$ & &\multicolumn{4}{c}{$-2.0$} & &... &$-2.4^{+1.5}_{-1.6}$ &$-2.6^{+1.4}_{-1.5}$ & &... &$-2.5^{+1.5}_{-1.6}$ &$-3.3^{+1.2}_{-1.1}$ & &... &$-2.4^{+1.6}_{-1.6}$ &$-2.9^{+1.3}_{-1.3}$ & &... &$-2.5^{+1.6}_{-1.6}$ &$-2.8^{+1.5}_{-1.4}$ \\
    &$f_\mathrm{C}$ & &\multicolumn{4}{c}{$0.7$} & &... &... &$0.4^{+0.2}_{-0.2}$ & &... &... &$0.8^{+0.1}_{-0.1}$ & &... &... &$0.6^{+0.2}_{-0.3}$ & &... &... &$0.4^{+0.4}_{-0.2}$ \\
    &$T_\mathrm{eq}\left[\mathrm{K}\right]$ & &\multicolumn{4}{c}{$249$} & &$256^{+9}_{-8}$ &$259^{+7}_{-7}$ &$251^{+9}_{-8}$ & &$260^{+5}_{-5}$ &$260^{+4}_{-5}$ &$253^{+5}_{-6}$ & &$259^{+6}_{-6}$ &$259^{+6}_{-5}$ &$252^{+6}_{-7}$ & &$261^{+3}_{-3}$ &$260^{+3}_{-3}$ &$256^{+4}_{-5}$ \\
    &$A_\mathrm{B}$ & &\multicolumn{4}{c}{$0.36$} & &$0.28^{+0.09}_{-0.10}$ &$0.25^{+0.08}_{-0.09}$ &$0.34^{+0.08}_{-0.10}$ & &$0.23^{+0.06}_{-0.06}$ &$0.24^{+0.05}_{-0.05}$ &$0.32^{+0.06}_{-0.06}$ & &$0.25^{+0.07}_{-0.08}$ &$0.25^{+0.06}_{-0.07}$ &$0.32^{+0.07}_{-0.07}$ & &$0.23^{+0.04}_{-0.04}$ &$0.24^{+0.04}_{-0.04}$ &$0.29^{+0.05}_{-0.05}$
\enddata
\addtolength{\leftskip} {-0cm}
\addtolength{\rightskip}{-7.3cm}
\tablecomments{Here, $L(\cdot)$ stands for $\lgrt{\cdot}$. In columns two to five, we list the ground truth values for the view. If independent of the atmospheric pressure, we provide a single value. Otherwise, we provide ground truth values at $1$~bar, $10^{-1}$~bar, $10^{-2}$~bar, and $10^{-3}$~bar where available. In the last twelve columns, we list the median and the $16\% - 84\%$ range (via $+/-$ indices) of the parameter posteriors for all combinations of \R{} (50, 100), \SN{} (10, 20), and retrieval model (\cwcf{}, \vwcf{}, \vwcl{}). For \ce{H2O}, all listed posterior values correspond to the abundances at the pressure \Popaque{} where the atmosphere becomes optically thick.}
\label{tab:Ret_SP}
\end{deluxetable}
\end{rotatetable}

\movetabledown=68mm
\begin{rotatetable}
\begin{deluxetable}{ccccccccccccccccccccccc}
\tablecaption{Numeric values for the EqC view posteriors shown in Figures~\ref{fig:views_posteriors_SN10} and \ref{fig:views_posteriors_SN20}.}
\tablehead{
&&&\multicolumn{4}{c}{Ground Truths} &&\multicolumn{7}{c}{Posteriors for \Rv{50} Spectra} &&\multicolumn{7}{c}{Posteriors for \Rv{100} Spectra} \\ \cline{9-15} \cline{17-23}
&&&\multicolumn{4}{c}{Pressure Levels [bar]} &&\multicolumn{3}{c}{\lifesim{} \SNv{10}}&&\multicolumn{3}{c}{\lifesim{} \SNv{20}} &&\multicolumn{3}{c}{\lifesim{} \SNv{10}}&&\multicolumn{3}{c}{\lifesim{} \SNv{20}} \\\cline{4-7} \cline{9-11} \cline{13-15} \cline{17-19} \cline{21-23}
&\colhead{Parameter} &&\colhead{1}&\colhead{$10^{-1}$}&\colhead{$10^{-2}$}&\colhead{$10^{-3}$} &&\cwcf{}&\vwcf{}&\vwcl{}&&\cwcf{}&\vwcf{}&\vwcl{}&&\cwcf{}&\vwcf{}&\vwcl{}&&\cwcf{}&\vwcf{}&\vwcl{}
}
\startdata
\multirow{13}{*}{\rotatebox[origin=c]{90}{EqC Jan View}}
    &$L\left(P_0\left[\mathrm{bar}\right]\right)$ & &\multicolumn{4}{c}{$0.0$} & &$-0.6^{+0.4}_{-0.4}$ &$0.5^{+0.9}_{-0.7}$ &$0.7^{+0.7}_{-0.8}$ & &$-0.4^{+0.4}_{-0.3}$ &$-0.3^{+0.4}_{-0.2}$ &$0.2^{+1.1}_{-0.6}$ & &$-0.5^{+0.3}_{-0.3}$ &$-0.5^{+0.3}_{-0.2}$ &$0.7^{+0.8}_{-0.7}$ & &$-0.4^{+0.3}_{-0.2}$ &$-0.4^{+0.2}_{-0.2}$ &$0.9^{+0.7}_{-0.7}$ \\
    &$L\left(P_\mathrm{Op.}\left[\mathrm{bar}\right]\right)$ & &\multicolumn{4}{c}{$0.0$} & &$-0.6^{+0.4}_{-0.4}$ &$-0.0^{+0.4}_{-0.4}$ &$0.1^{+0.3}_{-0.4}$ & &$-0.4^{+0.4}_{-0.3}$ &$-0.4^{+0.3}_{-0.2}$ &$-0.1^{+0.4}_{-0.3}$ & &$-0.5^{+0.3}_{-0.3}$ &$-0.5^{+0.3}_{-0.2}$ &$-0.1^{+0.3}_{-0.3}$ & &$-0.4^{+0.3}_{-0.2}$ &$-0.4^{+0.2}_{-0.2}$ &$0.0^{+0.3}_{-0.3}$ \\
    &$T_0\left[\mathrm{K}\right]$ & &\multicolumn{4}{c}{$280$} & &$281^{+10}_{-9}$ &$331^{+303}_{-46}$ &$408^{+331}_{-109}$ & &$285^{+6}_{-6}$ &$284^{+5}_{-4}$ &$320^{+381}_{-31}$ & &$284^{+7}_{-7}$ &$283^{+7}_{-6}$ &$451^{+431}_{-143}$ & &$285^{+4}_{-4}$ &$284^{+4}_{-3}$ &$495^{+486}_{-170}$ \\
    &$T_\mathrm{Op.}\left[\mathrm{K}\right]$ & &\multicolumn{4}{c}{$280$} & &$281^{+10}_{-9}$ &$284^{+10}_{-7}$ &$289^{+10}_{-8}$ & &$285^{+6}_{-6}$ &$284^{+5}_{-4}$ &$291^{+9}_{-7}$ & &$284^{+7}_{-7}$ &$283^{+6}_{-6}$ &$289^{+10}_{-7}$ & &$285^{+4}_{-4}$ &$284^{+4}_{-3}$ &$293^{+8}_{-7}$ \\
    &$R_\mathrm{pl}\left[R_\oplus\right]$ & &\multicolumn{4}{c}{$1.0$} & &$0.9^{+0.1}_{-0.1}$ &$0.9^{+0.1}_{-0.1}$ &$1.0^{+0.1}_{-0.1}$ & &$0.9^{+0.0}_{-0.0}$ &$0.9^{+0.0}_{-0.0}$ &$1.0^{+0.1}_{-0.1}$ & &$0.9^{+0.1}_{-0.1}$ &$0.9^{+0.0}_{-0.0}$ &$1.0^{+0.1}_{-0.1}$ & &$0.9^{+0.0}_{-0.0}$ &$0.9^{+0.0}_{-0.0}$ &$1.0^{+0.0}_{-0.0}$ \\
    &$L\left(\mathrm{CO_2}\right)$ & &\multicolumn{4}{c}{$-3.4$} & &$-2.4^{+0.9}_{-0.9}$ &$-3.7^{+0.9}_{-0.9}$ &$-3.5^{+0.7}_{-0.7}$ & &$-2.7^{+0.8}_{-0.7}$ &$-2.9^{+0.4}_{-0.5}$ &$-3.1^{+0.6}_{-0.6}$ & &$-2.6^{+0.9}_{-0.7}$ &$-2.7^{+0.6}_{-0.5}$ &$-3.3^{+0.6}_{-0.6}$ & &$-2.8^{+0.5}_{-0.6}$ &$-2.8^{+0.3}_{-0.3}$ &$-3.3^{+0.4}_{-0.4}$ \\
    &$L\left(\mathrm{O_3}\right)$ & &$-7.4$ &$-5.9$ &$-5.0$ &$-5.4$ & &$-5.8^{+0.7}_{-0.7}$ &$-6.6^{+0.5}_{-0.5}$ &$-6.7^{+0.5}_{-0.4}$ & &$-6.0^{+0.6}_{-0.5}$ &$-6.2^{+0.3}_{-0.4}$ &$-6.4^{+0.5}_{-0.5}$ & &$-5.9^{+0.7}_{-0.6}$ &$-6.0^{+0.4}_{-0.4}$ &$-6.6^{+0.4}_{-0.4}$ & &$-6.0^{+0.5}_{-0.5}$ &$-6.1^{+0.3}_{-0.3}$ &$-6.5^{+0.2}_{-0.2}$ \\
    &$L\left(\mathrm{CH_4}\right)$ & &$-6.0$ &$-6.1$ &$-6.4$ &$-6.6$ & &$-4.9^{+1.2}_{-1.6}$ &$-7.4^{+1.6}_{-1.5}$ &$-7.5^{+1.5}_{-1.5}$ & &$-4.8^{+0.8}_{-0.7}$ &$-5.2^{+0.5}_{-0.8}$ &$-5.9^{+1.0}_{-1.3}$ & &$-4.8^{+0.8}_{-0.7}$ &$-4.9^{+0.6}_{-0.7}$ &$-6.3^{+0.9}_{-1.4}$ & &$-4.8^{+0.5}_{-0.6}$ &$-5.0^{+0.4}_{-0.4}$ &$-5.9^{+0.4}_{-0.4}$ \\
    &$L\left(\mathrm{H_2O}\right)$ & &$-2.1$ &$-5.6$ &... &... & &$-2.5^{+0.7}_{-0.8}$ &$-2.1^{+0.4}_{-0.4}$ &$-2.1^{+0.3}_{-0.4}$ & &$-2.9^{+0.8}_{-0.7}$ &$-2.1^{+0.4}_{-0.5}$ &$-2.0^{+0.3}_{-0.4}$ & &$-2.8^{+0.8}_{-0.8}$ &$-2.0^{+0.4}_{-0.6}$ &$-1.9^{+0.3}_{-0.3}$ & &$-3.1^{+0.6}_{-0.6}$ &$-2.0^{+0.3}_{-0.5}$ &$-1.9^{+0.2}_{-0.2}$ \\
    &$L\left(\delta_\mathrm{H_2O}\right)$ & &\multicolumn{4}{c}{$-2.0$} & &... &$-2.6^{+1.5}_{-1.4}$ &$-2.6^{+1.4}_{-1.4}$ & &... &$-2.5^{+1.5}_{-1.4}$ &$-2.8^{+1.4}_{-1.3}$ & &... &$-2.5^{+1.5}_{-1.6}$ &$-3.1^{+1.4}_{-1.2}$ & &... &$-2.5^{+1.6}_{-1.5}$ &$-3.6^{+1.1}_{-1.0}$ \\
    &$f_\mathrm{C}$ & &\multicolumn{4}{c}{$0.7$} & &... &... &$0.5^{+0.2}_{-0.2}$ & &... &... &$0.5^{+0.2}_{-0.3}$ & &... &... &$0.6^{+0.1}_{-0.2}$ & &... &... &$0.7^{+0.1}_{-0.1}$ \\
    &$T_\mathrm{eq}\left[\mathrm{K}\right]$ & &\multicolumn{4}{c}{$257$} & &$264^{+10}_{-9}$ &$265^{+7}_{-6}$ &$255^{+9}_{-9}$ & &$268^{+5}_{-5}$ &$266^{+4}_{-4}$ &$259^{+6}_{-6}$ & &$267^{+7}_{-6}$ &$266^{+5}_{-5}$ &$257^{+7}_{-7}$ & &$269^{+3}_{-4}$ &$267^{+3}_{-3}$ &$261^{+5}_{-5}$ \\
    &$A_\mathrm{B}$ & &\multicolumn{4}{c}{$0.27$} & &$0.19^{+0.11}_{-0.12}$ &$0.17^{+0.08}_{-0.09}$ &$0.29^{+0.09}_{-0.10}$ & &$0.14^{+0.07}_{-0.07}$ &$0.17^{+0.05}_{-0.06}$ &$0.25^{+0.07}_{-0.07}$ & &$0.15^{+0.08}_{-0.09}$ &$0.17^{+0.07}_{-0.07}$ &$0.27^{+0.08}_{-0.08}$ & &$0.13^{+0.05}_{-0.05}$ &$0.15^{+0.04}_{-0.04}$ &$0.23^{+0.05}_{-0.06}$ \\\hline
\multirow{13}{*}{\rotatebox[origin=c]{90}{EqC Jul View}}
    &$L\left(P_0\left[\mathrm{bar}\right]\right)$ & &\multicolumn{4}{c}{$0.0$} & &$-0.5^{+0.4}_{-0.3}$ &$0.8^{+0.7}_{-0.7}$ &$0.6^{+0.8}_{-0.7}$ & &$-0.4^{+0.3}_{-0.2}$ &$-0.2^{+0.8}_{-0.3}$ &$0.8^{+0.7}_{-0.6}$ & &$-0.5^{+0.3}_{-0.3}$ &$-0.4^{+0.8}_{-0.3}$ &$0.8^{+0.7}_{-0.8}$ & &$-0.4^{+0.3}_{-0.2}$ &$-0.4^{+0.2}_{-0.1}$ &$0.4^{+1.0}_{-0.8}$ \\
    &$L\left(P_\mathrm{Op.}\left[\mathrm{bar}\right]\right)$ & &\multicolumn{4}{c}{$0.0$} & &$-0.5^{+0.4}_{-0.3}$ &$-0.1^{+0.4}_{-0.4}$ &$-0.0^{+0.4}_{-0.3}$ & &$-0.4^{+0.3}_{-0.2}$ &$-0.4^{+0.3}_{-0.3}$ &$0.1^{+0.3}_{-0.3}$ & &$-0.5^{+0.3}_{-0.3}$ &$-0.4^{+0.3}_{-0.2}$ &$-0.0^{+0.4}_{-0.3}$ & &$-0.4^{+0.3}_{-0.2}$ &$-0.4^{+0.2}_{-0.1}$ &$-0.1^{+0.3}_{-0.3}$ \\
    &$T_0\left[\mathrm{K}\right]$ & &\multicolumn{4}{c}{$285$} & &$285^{+11}_{-10}$ &$406^{+330}_{-110}$ &$389^{+387}_{-90}$ & &$288^{+6}_{-5}$ &$288^{+72}_{-6}$ &$474^{+368}_{-144}$ & &$287^{+8}_{-7}$ &$288^{+50}_{-7}$ &$468^{+435}_{-155}$ & &$289^{+4}_{-4}$ &$287^{+4}_{-3}$ &$381^{+411}_{-85}$ \\
    &$T_\mathrm{Op.}\left[\mathrm{K}\right]$ & &\multicolumn{4}{c}{$285$} & &$285^{+11}_{-10}$ &$284^{+11}_{-8}$ &$289^{+12}_{-8}$ & &$288^{+6}_{-5}$ &$285^{+7}_{-7}$ &$290^{+8}_{-6}$ & &$287^{+8}_{-7}$ &$285^{+7}_{-7}$ &$290^{+9}_{-7}$ & &$289^{+4}_{-4}$ &$287^{+4}_{-3}$ &$294^{+9}_{-7}$ \\
    &$R_\mathrm{pl}\left[R_\oplus\right]$ & &\multicolumn{4}{c}{$1.0$} & &$0.9^{+0.1}_{-0.1}$ &$1.0^{+0.1}_{-0.1}$ &$1.0^{+0.1}_{-0.1}$ & &$0.9^{+0.0}_{-0.0}$ &$1.0^{+0.1}_{-0.0}$ &$1.0^{+0.1}_{-0.1}$ & &$0.9^{+0.1}_{-0.1}$ &$1.0^{+0.1}_{-0.0}$ &$1.0^{+0.1}_{-0.1}$ & &$0.9^{+0.0}_{-0.0}$ &$0.9^{+0.0}_{-0.0}$ &$1.0^{+0.0}_{-0.0}$ \\
    &$L\left(\mathrm{CO_2}\right)$ & &\multicolumn{4}{c}{$-3.4$} & &$-2.5^{+0.9}_{-0.8}$ &$-3.5^{+0.8}_{-0.9}$ &$-3.5^{+0.7}_{-0.7}$ & &$-2.7^{+0.7}_{-0.7}$ &$-2.6^{+0.6}_{-0.6}$ &$-3.4^{+0.5}_{-0.5}$ & &$-2.6^{+0.8}_{-0.7}$ &$-2.6^{+0.5}_{-0.5}$ &$-3.3^{+0.6}_{-0.5}$ & &$-2.9^{+0.6}_{-0.6}$ &$-2.7^{+0.3}_{-0.3}$ &$-3.0^{+0.5}_{-0.5}$ \\
    &$L\left(\mathrm{O_3}\right)$ & &$-7.3$ &$-6.0$ &$-5.0$ &$-5.4$ & &$-5.8^{+0.7}_{-0.6}$ &$-6.5^{+0.5}_{-0.5}$ &$-6.6^{+0.5}_{-0.4}$ & &$-5.9^{+0.6}_{-0.5}$ &$-6.0^{+0.4}_{-0.4}$ &$-6.6^{+0.3}_{-0.3}$ & &$-5.9^{+0.7}_{-0.5}$ &$-5.9^{+0.4}_{-0.4}$ &$-6.5^{+0.4}_{-0.4}$ & &$-6.0^{+0.5}_{-0.5}$ &$-5.9^{+0.3}_{-0.3}$ &$-6.3^{+0.4}_{-0.3}$ \\
    &$L\left(\mathrm{CH_4}\right)$ & &$-6.0$ &$-6.1$ &$-6.4$ &$-6.6$ & &$-5.0^{+1.1}_{-1.6}$ &$-7.4^{+1.6}_{-1.6}$ &$-7.4^{+1.5}_{-1.5}$ & &$-4.8^{+0.7}_{-0.7}$ &$-5.1^{+0.7}_{-1.1}$ &$-6.3^{+0.7}_{-1.0}$ & &$-4.8^{+0.8}_{-0.7}$ &$-5.0^{+0.6}_{-1.2}$ &$-6.3^{+0.8}_{-1.4}$ & &$-4.9^{+0.6}_{-0.6}$ &$-4.9^{+0.4}_{-0.4}$ &$-5.6^{+0.7}_{-0.5}$ \\
    &$L\left(\mathrm{H_2O}\right)$ & &$-2.1$ &$-5.6$ &... &... & &$-2.6^{+0.8}_{-0.8}$ &$-2.1^{+0.3}_{-0.4}$ &$-2.0^{+0.3}_{-0.4}$ & &$-3.0^{+0.7}_{-0.7}$ &$-1.9^{+0.3}_{-0.5}$ &$-2.0^{+0.3}_{-0.3}$ & &$-2.8^{+0.8}_{-0.7}$ &$-1.9^{+0.4}_{-0.6}$ &$-1.9^{+0.3}_{-0.3}$ & &$-3.1^{+0.6}_{-0.6}$ &$-1.9^{+0.3}_{-0.6}$ &$-1.8^{+0.3}_{-0.3}$ \\
    &$L\left(\delta_\mathrm{H_2O}\right)$ & &\multicolumn{4}{c}{$-2.0$} & &... &$-2.6^{+1.6}_{-1.4}$ &$-2.7^{+1.4}_{-1.4}$ & &... &$-2.6^{+1.5}_{-1.5}$ &$-3.4^{+1.2}_{-1.0}$ & &... &$-2.5^{+1.5}_{-1.5}$ &$-3.1^{+1.4}_{-1.2}$ & &... &$-2.6^{+1.6}_{-1.5}$ &$-3.5^{+1.1}_{-0.9}$ \\
    &$f_\mathrm{C}$ & &\multicolumn{4}{c}{$0.7$} & &... &... &$0.4^{+0.2}_{-0.2}$ & &... &... &$0.6^{+0.1}_{-0.2}$ & &... &... &$0.5^{+0.1}_{-0.2}$ & &... &... &$0.6^{+0.2}_{-0.3}$ \\
    &$T_\mathrm{eq}\left[\mathrm{K}\right]$ & &\multicolumn{4}{c}{$261$} & &$266^{+10}_{-10}$ &$264^{+7}_{-6}$ &$257^{+9}_{-9}$ & &$270^{+5}_{-5}$ &$264^{+5}_{-6}$ &$260^{+6}_{-6}$ & &$270^{+7}_{-7}$ &$266^{+6}_{-6}$ &$258^{+6}_{-7}$ & &$271^{+4}_{-3}$ &$268^{+3}_{-3}$ &$262^{+5}_{-5}$ \\
    &$A_\mathrm{B}$ & &\multicolumn{4}{c}{$0.23$} & &$0.16^{+0.12}_{-0.14}$ &$0.19^{+0.07}_{-0.09}$ &$0.27^{+0.09}_{-0.11}$ & &$0.11^{+0.07}_{-0.07}$ &$0.19^{+0.07}_{-0.07}$ &$0.24^{+0.07}_{-0.07}$ & &$0.12^{+0.09}_{-0.10}$ &$0.17^{+0.07}_{-0.07}$ &$0.26^{+0.08}_{-0.08}$ & &$0.10^{+0.05}_{-0.05}$ &$0.13^{+0.04}_{-0.05}$ &$0.22^{+0.06}_{-0.06}$
\enddata
\addtolength{\leftskip} {0cm}
\addtolength{\rightskip}{-7.3cm}
\tablecomments{Here, $L(\cdot)$ stands for $\lgrt{\cdot}$. In columns two to five, we list the ground truth values for the view. If independent of the atmospheric pressure, we provide a single value. Otherwise, we provide ground truth values at $1$~bar, $10^{-1}$~bar, $10^{-2}$~bar, and $10^{-3}$~bar where available. In the last twelve columns, we list the median and the $16\% - 84\%$ range (via $+/-$ indices) of the parameter posteriors for all combinations of \R{} (50, 100), \SN{} (10, 20), and retrieval model (\cwcf{}, \vwcf{}, \vwcl{}). For \ce{H2O}, all listed posterior values correspond to the abundances at the pressure \Popaque{} where the atmosphere becomes optically thick.}
\label{tab:Ret_EqC}
\end{deluxetable}
\end{rotatetable}

\end{document}